\documentclass{article}
\usepackage{xcolor}

\usepackage{graphicx} 
\usepackage{hyperref}
\usepackage{algorithm}
\usepackage{algorithmic}
\usepackage{amsmath}
\usepackage{amssymb}
\usepackage[capitalise]{cleveref}
\usepackage{booktabs}
\usepackage{multirow}
\usepackage{authblk}
\usepackage{subcaption}
\usepackage{xr}
\usepackage{multibib}  
\newcites{SI}{References for the Supplementary Information} 

\title{GPU-Accelerated Charge-Equilibration for Shadow Molecular Dynamics in Python}

\author{
    Mehmet Cagri Kaymak\textsuperscript{1}, 
    Nicholas Lubbers\textsuperscript{1},
    Christian F. A. Negre\textsuperscript{1}, 
    Michael Wall\textsuperscript{1}, 
    Anders M.N. Niklasson\textsuperscript{1} \\
    \textsuperscript{1}Los Alamos National Laboratory, Los Alamos, NM, USA
}
\date{}
\begin{document}

\maketitle
\begin{abstract}
With recent advancements in machine learning for interatomic potentials, Python has become the go-to programming language for exploring new ideas. While machine-learning potentials are often developed in Python-based frameworks, existing molecular dynamics software is predominantly written in lower-level languages. This disparity complicates the integration of machine learning potentials into these molecular dynamics libraries. Additionally, machine learning potentials typically focus on local features, often neglecting long-range electrostatics due to computational complexities. This is a key limitation as applications can require long-range electrostatics and even flexible charges to achieve the desired accuracy. Recent charge equilibration models can address these issues, but they require iterative solvers to assign relaxed flexible charges to the atoms. Conventional implementations also demand very tight convergence to achieve long-term stability, further increasing computational cost. In this work, we present a scalable Python implementation of a recently proposed shadow molecular dynamics scheme based on a charge equilibration model, which avoids the convergence problem while maintaining long-term energy stability and accuracy of observable properties. To deliver a functional and user-friendly Python-based library, we implemented an efficient neighbor list algorithm, Particle Mesh Ewald, and traditional Ewald summation techniques, leveraging the GPU-accelerated power of Triton and PyTorch. We integrated these approaches with the Python-based shadow molecular dynamics scheme, enabling fast charge equilibration for scalable machine learning potentials involving systems with hundreds of thousands of atoms.
\end{abstract}
 
\section{Introduction}
Molecular dynamics (MD) is a widely used method to simulate and understand phenomena at the atomistic scale in chemistry, molecular biology, and materials science.  The accuracy of MD simulations depends on how well one can capture the interactions between the atoms. While first-principles approaches such as density functional theory \cite{hohen,KohnSham65,RParr89,ROJones89,RMDreizler90,JPerdew92,Becke93, WKohn99} provide highly accurate results, their computational complexity typically limits their applicability in molecular dynamics simulations to systems with at most a few hundred atoms. In contrast, traditional classical force field models can scale to systems with millions of atoms thanks to simplifying assumptions, such as constrained bond connectivity and fixed charges. However, these assumptions limit their ability to capture the dynamics of, for example, chemical reactions, flexible charge transfer, and polarization \cite{monticelli2013force, senftle2016reaxff}.

Machine Learning (ML) has been able to improve the performance of classical force field models by allowing highly flexible functional forms for the local atom-projected energies and forces using large reference datasets for training. While these Machine Learning Potentials (MLPs) successfully capture the local environment of atoms, they struggle with long-range effects when there is no physical inductive bias in their functional form, largely due to the curse of dimensionality. In recent years, MLPs combined with various forms of charge equilibration (QEQ) models \cite{FJVesely77,WJMortier86,MSprik88,MSprik90,AKRappe91,DVanBelle92,WSRick94,TAHalgren01,GLamoureaux03,GAKaminsky04,PEMLopes09,PCieplak09,SNaserifar17,JZhifeng19} have shown promise in capturing long-range electrostatic interactions \cite{SGoedecker15,TWKo20,TWKo23}. Although including a physically grounded long-range electrostatic interaction improves chemical fidelity, it requires solving a linear system of equations to determine the relaxed charges that minimize the system's electrostatic energy (or finding the corresponing electronegativity equilibration) prior to the force evaluation in MD simulations, which increases the computational complexity, especially for large systems.

Iterative linear solvers can alleviate some of these challenges, but they introduce convergence and stability problems. Energy conservation in MD simulations with QEQ models is highly sensitive to the solver's convergence tolerance \cite{nomura2015extended}. Tighter tolerances improve stability but also increase computational cost, as they require multiple Coulomb summations, for example, using repeated Ewald summations for systems with periodic boundary conditions. A full construction of the Coulomb potential is required in each iteration. Additionally, while MLPs are predominantly developed in Python due to the availability of machine learning frameworks such as PyTorch \cite{paszke2019pytorch}, there is a lack of efficient Python implementations, not only of the essential and often time-limiting components of various forms of the Ewald summation, but also of scalable neighbor list generation. MD libraries like LAMMPS have highly optimized implementations for these components in C/C++, but integrating them tightly with Python-based MLPs slows down the development, especially when dealing with fluctuating charge models.

In this work, we present highly efficient implementations of neighbor list generation, regular Ewald summation, and Particle Mesh Ewald (PME) algorithms in Python, leveraging PyTorch and the Triton compiler \cite{tillet2019triton} for high-performance GPU acceleration. We also introduce the first scalable implementation of shadow molecular dynamics based on charge equilibration in combination with extended Lagrangian Born-Oppenheimer molecular dynamics (XL-BOMD), which relaxes the convergence criteria of the iterative solver while maintaining long-term energy stability and accuracy of MD observables \cite{ANiklasson21,JGoff23,Niklasson23}. Our approach combines innovative physics reformulations with fast, scalable solvers, enabling robust MD simulations on high-performance hardware.

\section{Related Work}

Including flexible charges is essential for modeling reactivity in molecular systems. The Reactive Force Field (ReaxFF) employs the QEQ method to assign dynamic charges to atoms, relying on a tapered approximation for electrostatic interactions \cite{senftle2016reaxff}. In this approach, electrostatic interactions smoothly diminish to zero beyond a cutoff distance, typically around 10 \AA. This approximation avoids the computationally intensive Ewald summation, simplifying charge equilibration as the resulting Coulomb matrix becomes sparse for sufficiently large systems.

Efficient implementations of ReaxFF and QEQ models are available in widely used MD libraries, such as LAMMPS \cite{thompson2022lammps}, PuReMD \cite{aktulga2012parallel}, and JAX-MD \cite{kaymak2023end}. JAX-MD additionally offers fast neighbor list calculations and supports regular and Particle Mesh Ewald implementations in Python, though charge equilibration is currently only available for tapered Coulomb interactions.

Scalable QEQ implementations using the Particle Mesh Ewald algorithm have also been explored; however, only CPU-based solutions have been presented \cite{gubler2024accelerating}. In this work, we provide a hardware-portable Python implementation leveraging PyTorch and Triton, along with a scalable neighbor list implementation, which is essential for the real-space part of the Ewald sum. Further details are provided in the next section.

Finally, shadow molecular dynamics for flexible charge models was recently introduced to improve the stability and to accelerate QEQ-based MD simulations \cite{ANiklasson21,ANiklasson21b,JGoff23,CHLi2025}. However, the existing work is so far primarily focused on the underlying theory or the ML optimization using the atomic cluster expansion. The demonstrations are mainly proof-of-concept simulations of smaller systems with at most a few hundred atoms. In this work, we offer a scalable Python implementation for shadow MD based on charge equilibration models that is suitable also for very large systems, including hundreds of thousands of atoms. The application and demonstration of this recent approach to large-scale MD simulations is an important step towards affordable, physically faithful simulations. 

The shadow molecular dynamics scheme for flexible charge models provides a more consistent and theoretically flexible formulation over previous XL-BOMD schemes for charge equilibration models \cite{KNomura15,AAlbaugh15,AAlbaugh18,ILeven19}. However, the underlying framework of XL-BOMD for propagating the extended electronic degrees of freedom, which originally was developed for orbital-resolved first principles electronic structure theory, is the same \cite{ANiklasson06,ANiklasson08,ANiklasson21b}.

\section{Design and Implementation}

In this section, we review the mathematics underpinning our XL-BOMB charge equilibration code, with special attention to algorithmic considerations. First, we describe the neighbors list implementations, followed by two forms of Ewald summation algorithms, and finally, the algorithms for the dynamics, both without and with XL-BOMD.

\subsection{Cell based Neighbor List}

\begin{figure}
\centering
        \includegraphics[totalheight=6cm]{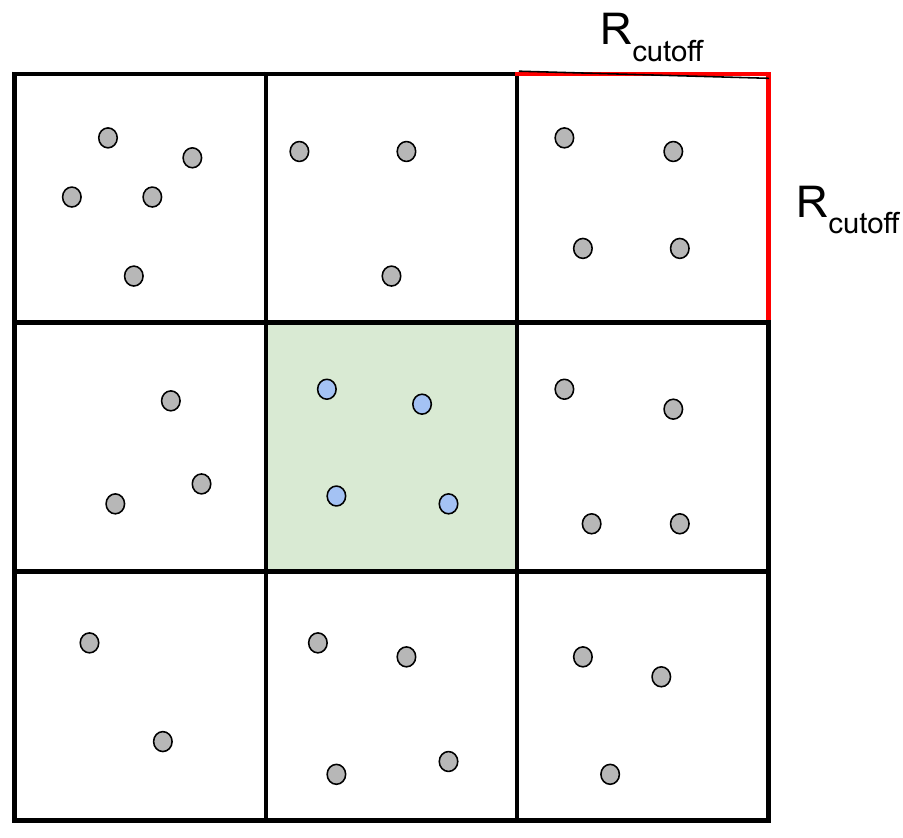}
    \caption{Illustration of the cell based neighbor search method.}
    \label{fig:cell_neighbor}
\end{figure}
The cell-based neighbor list algorithm is a highly efficient method for managing pairwise short-range interactions in large particle systems \cite{MAllen90}. By leveraging spatial decomposition, it reduces computational complexity by focusing on local interactions. In this method, the simulation domain is divided into a grid of cells, with each cell containing particles located within its bounds. Typically, the grid size is chosen to match or slightly exceed the cutoff distance of the pairwise interactions, simplifying the neighbor search process. For any given atom, we only need to search neighboring cells to identify nearby atoms within the cutoff range, reducing the number of unnecessary comparisons. Periodic boundary conditions are easily incorporated by using the minimum image convention. A 2-dimensional example of this process is shown in \cref{fig:cell_neighbor}. Leveraging PyTorch’s tensor operations, the neighbor list can be constructed and queried in a fully vectorized fashion, minimizing the need for explicit loops. This approach reduces computational complexity and maximizes the benefits of parallel processing, making it highly suitable for high-dimensional data and large-scale simulations. A similar implementation to our approach has been provided for JAX \cite{schoenholz2020jax}.

We implement two output formats for the neighbor list: the ELLPACK format and the Coordinate List (COO) format. In the ELLPACK format, the neighbors for each particle are stored in a dense matrix of size \(N \times K\), where \(N\) is the number of particles and \(K\) is the maximum number of neighbors per particle. This format is computationally efficient for subsequent pairwise interaction calculations but requires allocating memory for the maximum possible number of neighbors, which can increase with particle density or cutoff distance. The COO format uses a sparse representation, storing only the valid neighbor pairs, which reduces memory usage but may introduce overhead when accessing the neighbor data \cite{filippone2017sparse}.

One challenge with the vectorized approach is that it requires gathering all candidate neighbors together. For the 3-D case, this means collecting \(3^3 \times C_{\text{max}}\) elements for each atom, where $C_{\text{max}}$ is the maximum number of particles per cell, to fully vectorize the filtering operation, as shown in \cref{algo:nbr_list}. This approach leads to high memory usage, particularly for large dense systems. A more memory-efficient alternative is to iterate through neighboring cells one by one, applying filtering individually. Since explicit loops are suboptimal in PyTorch, we use a Triton kernel for this operation, allowing the looping over cells to happen within the Triton kernel. This approach achieves up to 2$\times$ speedup on modern GPUs, as shown in \cref{fig:neighbor_list_perf}. The general outline of our vectorized PyTorch neighbor list algorithm is given in \cref{algo:nbr_list}.

In the Triton implementation, we launch a kernel where each instance processes a single particle, iterating through neighboring cells and applying filtering in a block by block fashion. This process involves allocating an intermediate tensor to store the filtered results, with the size based on an estimated number of neighbors per atom. Despite this additional allocation, the method has lower memory requirements overall, as only the filtered results are stored. To optimize memory usage, we estimate the number of neighbors, incorporating a buffer to accommodate variations and enable the same estimate to be reused across multiple simulation steps.

\begin{algorithm}
\caption{\textsc{Generate Neighbor List}}
\begin{algorithmic}[1]
\STATE $cells : [n_x, n_y, n_z, C_{max}]$ tensor where $n_x$, $n_y$, $n_z$ are the cells per dimension $X$, $Y$, $Z$, and $C_{max}$ is the maximum number of particles per cell.
\STATE $p\_map : [3,N]$ tensor where $N$ is the number of atoms. It contains the mapping of particles to cell indices.
\STATE $shifts :$ list containing 26 possible neighboring cell shifts based on the minimum image convention in 3-D space.
\STATE Initialize $candidates \leftarrow [cells]$
\FOR{each $shift$ in $shifts$}
    \STATE Compute the neighboring cells for $shift$ and append them to $candidates$.
\ENDFOR
\STATE Concatenate $candidates$ for vectorized processing, $[n_x, n_y, n_z, 27, C_{max}]$.
\STATE $candidates \leftarrow candidates[p\_map[0], p\_map[1], p\_map[2]]$ // Select the candidates of each atom, results in $[N, 27, C_{max}]$ tensor.
\STATE $neighbors \leftarrow \text{filter}(candidates)$ // Filter the candidates based on the cutoff distance and construct the neighbor list tensor.
\RETURN $neighbors$
\end{algorithmic}
\label{algo:nbr_list}
\end{algorithm}

\subsection{Regular Ewald Summation}
The regular Ewald summation is a widely used method for calculating long-range interactions in periodic systems, especially for electrostatic interactions. It works by splitting the potential into two parts: a screened short-range interactions term that decays quickly in real space and a smooth long-range term that can be handled in reciprocal space. In the real-space part, interactions are computed within a finite cutoff radius, simplifying the calculations while retaining accuracy. The reciprocal-space part involves a Fourier transform, efficiently capturing the contribution of distant interactions. By balancing these two terms, Ewald summation achieves time complexity of $O(N^{3/2})$, where $N$ is the number of atoms in the cell with periodic boundary conditions \cite{toukmaji1996ewald}. It is quite remarkable that making an all-to-all interacting system infinite in size by introducing periodic boundary conditions can, in fact, reduce the computational cost. 

The Coulomb energy of a system can be expressed as
\begin{equation}
E_{\text{Coul}} = \frac{1}{2} \sum_{i=1}^{N} \sum_{j \neq i} q_i q_j \phi(\mathbf{r}_{ij}),
\label{eq:ewald_general}
\end{equation}
where $q_i$ is the partial charge of particle $i$, and \( \phi(\mathbf{r}_{ij}) \) is the electrostatic $\propto 1/\vert {\bf r}_{ij}\vert$ potential at a distance \( \mathbf{r}_{ij} = \mathbf{r}_i - \mathbf{r}_j \). The summation in \cref{eq:ewald_general} is not necessarily convergent when we have periodic boundary condition because each atom has infinite number of neighbors.
As mentioned, the basic idea behind the Ewald summation is to split the summation. We can do the splitting in three terms, 
\begin{equation}
   E_{\text{Coul}} = E_{\text{real}} + E_{\text{rec}} + E_{\text{self}}
\end{equation}
The splitting of $E_{\text{Coul}}$ into the real ($E_{\text{real}}$), reciprocal ($E_{\text{rec}}$) and a self energy term ($E_{\text{self}}$) components make the summation conditionally convergent for charge neutral systems \cite{JZiman79,MAllen90}. The self-energy term, $E_{\text{self}}$, is used to correct for the artificial interaction of each particle with its own periodic images.
The screened real-space term is given by
\begin{equation}
E_{\text{real}} = \frac{1}{2} \sum_{i=1}^{N} \sum_{j \neq i} q_i q_j \sum_{\mathbf{n}} \frac{\operatorname{erfc}(\alpha |\mathbf{r}_{ij} + \mathbf{n}L|)}{|\mathbf{r}_{ij} + \mathbf{n}L|},
\label{eq:ewald_real}
\end{equation}

where \( \mathbf{n} \) is a vector of integers representing periodic images, \( L \) is the length of the unit cell, \( \operatorname{erfc} \) is the complementary error function, and \( \alpha \) is a convergence parameter that controls the division between real and reciprocal space terms. For non-cubic unit cell, we may generalize the box length to general lattice vectors. The reciprocal long-range term is given by
\begin{equation}
E_{\text{rec}} = \frac{1}{2V} \sum_{\mathbf{k} \neq 0} \frac{4 \pi}{|\mathbf{k}|^2} e^{-\frac{|\mathbf{k}|^2}{4\alpha^2}} \left| \sum_{j=1}^{N} q_j e^{i \mathbf{k} \cdot \mathbf{r}_j} \right|^2,
\label{eq:ewald_rec}
\end{equation}
where \( V \) is the volume of the simulation cell, \( \mathbf{k} \) is the reciprocal lattice vector, and the exponential factor \( e^{-\frac{|\mathbf{k}|^2}{4\alpha^2}} \) ensures convergence.
Finally the additional self-energy term is
\begin{equation}
   E_{\text{self}} = -\frac{\alpha}{\sqrt{\pi}} \sum_{i=1}^{N} q_i^2
   \label{eq:ewald_self}.
\end{equation}

For the calculation of $E_{\text{real}}$, we utilize an ELLPACK-like sparse matrix neighbor list format, where the neighbors within a certain cutoff are stored as an $N \times K$ dense matrix, where $N$ is the number of particles in the box and $K$ is the maximum number of neighbors \cite{young2014iterative}. This format greatly simplifies the summation in \cref{eq:ewald_real} as we can do the summation over neighbors of each atom to calculate forces and charge derivatives ($\partial E_{real}/\partial q$) without using any atomic operation in PyTorch. While these operations prevent race conditions, they also introduce computational overhead due to synchronization. We also provide a Triton based implementation that improves the performance.

For the summation of the long-range reciprocal part, \(E_{\text{rec}}\), we store the reciprocal lattice vectors \(\{{\bf k}\}\) in the matrix \(\mathbf{K} \in \mathbb{R}^{k \times 3}\), where \(k\) is the total number of reciprocal vectors used. Additionally, we precalculate all the $\frac{4 \pi}{|\mathbf{k}|^2} e^{-\frac{|\mathbf{k}|^2}{4\alpha^2}}$ terms from \cref{eq:ewald_rec} and store them in a vector,  $\mathbf{v}$. This structured representation simplifies the computation in PyTorch, allowing us to leverage efficient linear algebra operations instead of explicit loops, improving performance.

Building on this, we can rewrite \cref{eq:ewald_rec} as
\begin{equation}
E_{\text{rec}} = \frac{2 \pi}{V} \sum_{j=1}^{k} v_j \left( s_j^{\text{Re}} s_j^{\text{Re}} + s_j^{\text{Im}} s_j^{\text{Im}} \right),
\end{equation}
where
\begin{equation}
s_j^{\text{Re}} = \sum_{i=1}^{N} S_{ij}^{\text{Re}}, \quad
s_j^{\text{Im}} = \sum_{i=1}^{N} S_{ij}^{\text{Im}}, \quad \text{for each } j = 1, \dots, k,
\end{equation}
and
\begin{equation}
S_{ij}^{\text{Re}} = \cos(M_{ij}) q_i, \quad
S_{ij}^{\text{Im}} = \sin(M_{ij}) q_i.
\label{eq:ewald_S}
\end{equation}
Here, \( \text{Re} \) and \( \text{Im} \) denote the real and imaginary parts, respectively. The matrix \( \mathbf{M} \) is defined element-wise as
\begin{equation}
M_{ij} = \mathbf{r}_i \cdot \mathbf{k}_j,
\label{eq:ewald_M}
\end{equation}
where \( \mathbf{r}_i \) represents the \( i \)th position vector in real space, and \( \mathbf{k}_j \) represents the \( j \)th reciprocal lattice vector, with both \( \mathbf{r}_i \) and \( \mathbf{k}_j \) $\in \mathbb{R}^3$. We can further express the force terms \( \frac{\partial E_{\text{rec}}}{\partial r_{i\alpha}} \), with subindex $\alpha = \{x,y,z\}$,  and the potential terms \( \frac{\partial E_{\text{rec}}}{\partial q_{i}} \) in more efficient forms:

\begin{equation}
\frac{\partial E_{\text{rec}}}{\partial r_{i\alpha}} =  \frac{4 \pi}{V} \sum_{j=1}^{k} K_{j\alpha} L_{ij},
\end{equation}

\begin{equation}
\frac{\partial E_{\text{rec}}}{\partial q_{i}} =  \frac{4 \pi}{V} \sum_{j=1}^{k} v_j \left( \cos(M_{ij}) s_j^{\text{Re}} + \sin(M_{ij}) s_j^{\text{Im}} \right),
\end{equation}

where
\begin{equation}
L_{ij} = v_j \left( S_{ij}^{\text{Re}} s_j^{\text{Re}} - S_{ij}^{\text{Im}} s_j^{\text{Im}} \right),
\end{equation}

This ensures that the phase terms in the summations are computed efficiently using matrix operations instead of loops.

As shown in \cref{eq:ewald_M}, while the calculation of \(E_{\text{rec}}\) and its derivatives can be fully vectorized for efficient computation in PyTorch, expressing it in this form requires a large intermediate matrix \(\mathbf{M}\) of size \(N \times k\). Although precomputing \(\mathbf{M}\) reduces computational complexity for derivative calculations, it significantly increases memory usage, limiting scalability. To address this, we provide a Triton implementation that computes \(\mathbf{M}\) in a block-wise manner on-the-fly using fast shared memory, avoiding global memory storage of \(\mathbf{M}\). This approach necessitates recomputing \cref{eq:ewald_S} and \cref{eq:ewald_M} in a block-wise fashion for the derivatives, but it improves both memory efficiency and performance on modern GPUs, as demonstrated in the next section. For a fair comparison, we also implement a Triton version of the fully vectorized approach described above, without memory optimizations. Finally, the accuracy can be adjusted using the Ewald accuracy parameter, which controls the number of reciprocal lattice vectors based on the desired accuracy. We utilize the same tuning heuristics as OpenMM \cite{eastman2023openmm}.

\subsection{Particle Mesh Ewald Summation}
The Particle Mesh Ewald (PME) method is an extension of the standard Ewald summation technique, aimed at efficiently calculating long-range electrostatic interactions in systems with periodic boundary conditions \cite{darden1993particle}. While the traditional Ewald summation relies on splitting the potential into a short range real-space  and long-range reciprocal-space components, the PME method enhances computational efficiency for large systems by relying on the Fast Fourier Transform (FFT) method \cite{JCooley65} for the reciprocal-space term (\cref{eq:ewald_rec}).
This leads to a reduction in the computational complexity from \(O(N^{3/2})\) to \(O(N \log N)\) \cite{toukmaji1996ewald,darden1993particle}, which can drastically improve performance on large enough systems. This is achieved by mapping the charges onto a discrete grid, known as the particle mesh, where the charge density is interpolated, here, using B-spline interpolation \cite{de1978practical}. The FFT is then applied to compute the reciprocal-space interactions efficiently on this grid. Once the reciprocal-space term is computed, the results are mapped back from the mesh to the particle positions. The PME approach allows simulations to handle much larger systems with periodic boundary conditions, while maintaining an accuracy comparable to the traditional Ewald summation. Similar to the regular Ewald summation, the PME accuracy can be tuned. The accuracy in the calculations of the reciprocal energy term increases or decreases the number of grid points and the order of the B-spline interpolation.  We utilize the same tuning heuristics as in OpenMM \cite{eastman2023openmm}.

\begin{algorithm}
\caption{Smooth PME Algorithm for $E_{\rm rec}$}
\begin{algorithmic}[1]
\STATE \textbf{Input:} Particle positions, charges, simulation box dimensions, Ewald parameter $\alpha$, grid dimensions, spline order
\STATE \textbf{Output:} Electrostatic energy for the periodic system

\STATE \textbf{Step 1: Mapping Charges to the Grid} \\
Map particle charges onto a 3D grid using cardinal B-spline interpolation, distributing each charge smoothly over nearby grid points.

\STATE \textbf{Step 2: Perform FFT on the Grid} \\
Execute a 3D Fast Fourier Transform (FFT) on the charge-mapped grid (real-to-complex), moving into reciprocal space.

\STATE \textbf{Step 3: Compute the total energy} \\
Scale the Fourier-transformed grid values with the Gaussian damping factors to filter out short-range contributions and sum the scaled norm of the complex grid values to obtain the total electrostatic energy ($E_{rec}$).

\STATE \textbf{Step 4: Inverse FFT to Real Space} \\
Apply an inverse FFT to transform the modified grid back to real space to compute forces and charge derivatives

\STATE \textbf{Step 5: Calculate the derivatives} \\
Calculate ${\partial E_{rec}}/{\partial r_{i}}$ and
${\partial E_{rec}}/{\partial q_{i}}$ by gathering partial derivatives (based on the window function used in Step-1).
\RETURN $E_{rec}$, ${\partial E_{rec}}/{\partial r_{i}}$ and
${\partial E_{rec}}/{\partial q_{i}}$ \text{for each } i = 1, \dots, N
\end{algorithmic}
\label{algo:PME}
\end{algorithm}
The details of the smooth PME algorithm for the calculation of the reciprocal energy term, $E_{\rm rec}$, are provided in \cref{algo:PME}. Our PME implementation is written entirely in PyTorch. For Step 1, we use atomic operations to distribute charge contributions over the grid points. The forward and inverse FFT operations are executed using PyTorch’s native functionality, which calls well-tuned, vendor-specific functions based on the hardware. Although scatter (Step 1) and gather (Step 5) operations could be further optimized by calculating B-spline values entirely in shared memory rather than accessing global memory, the dynamic shared memory indexing logic that is required for this is not yet supported by the Triton compiler. While hardware-specific compilers could enable these operations, we choose to keep everything in Python in a hardware-agnostic manner, making it easier to use across various hardware platforms. As demonstrated in \cref{subsec:performance}, the final implementation achieves high performance without compromising portability.

\subsection{Regular Born-Oppenheimer MD for a Charge Equilibration Model}
The charge equilibration (QEQ) model relies on a charge-dependent energy function, 
\begin{equation}
    E({\bf R,q}) =  \sum_i q_i \chi_i + \frac{1}{2} \sum_i q_i^2 u_i + E_{\text{Coul}}({\bf R}), \label{eq:qeq_energy_func}
\end{equation}
where ${\bf q} = \{q_i\}$ represents the net partial charge on each atom $i$, and $u_i$ and $\chi_i$ are the hardness and electronegativity parameters of atom $i$, respectively. $\mathbf{R}$ denotes the set of atomic positions, $(r_1, r_2, \ldots, r_N)$, with $r_i \in \mathbb{R}^3$. The term $E_{\text{Coul}}$ accounts for the Coulomb energy, as defined in \cref{eq:ewald_general}.

In Born-Oppenheimer (BO) MD based on the Born-Oppenheimer approximation \cite{WHeitler27,MBorn27,DMarx00,MTuckerman02}, interatomic forces are derived from the potential energy surface for the equilibrated ground-state solution of the electronic structure at fixed nuclear positions.  As a result, the energy $E({\bf R,q})$ in \cref{eq:qeq_energy_func} must be minimized with respect to the distribution of the partial charges, $\mathbf{q}$, under the condition that their sum is some given total net charge, $Q_{\rm tot}$. This constrained minimization give us the Born-Oppenheimer potential,
\begin{equation}
    U_{\rm BO}({\bf R}) = V_{\rm short}({\bf R}) + \min_{\bf q} \Big\{E({\bf R,q}) \Big \vert \sum_i q_i = Q_{\rm tot} \Big\}, \label{eq:BO_Min}
\end{equation}
where $V_{\rm short}({\bf R})$ is some charge-independent short-range potential. The total net charge constraint can be incorporated with the Lagrange multiplier method. The modified function, or Lagrangian, is expressed as 
\begin{equation}
    \mathcal{L}({\bf R,q},\lambda) = E({\bf R,q}) + \lambda \left( 
\sum_i q_i - Q_{\rm tot}\right).
\end{equation}
The solution to the constrained minimization problem is obtained by finding the stationary solution of $\mathcal{L}({\bf R,q},\lambda)$ with respect to both $\mathbf{q}$ and $\lambda$, i.e. we need to find a solution to
\begin{equation}
\frac{\partial \mathcal{L}}{\partial q_i} = \frac{\partial E}{\partial q_i} + \lambda = 0, \quad \forall i, \label{eq:qeq_partial_q}
\end{equation}
\begin{equation}
\frac{\partial \mathcal{L}}{\partial \lambda} = \sum_i q_i - Q_{\rm tot} = 0. \label{eq:qeq_partial_lambda}
\end{equation}
\cref{eq:qeq_partial_q} reflects the principle of electronegativity equalization \cite{sanderson1951interpretation}. We can formulate this algebraically as 
\begin{equation}
\begin{bmatrix}
\mathbf{C}& \mathbf{1} \\
\mathbf{1}^\top & 0
\end{bmatrix}
\begin{bmatrix}
\mathbf{q} \\
\lambda
\end{bmatrix}
=
\begin{bmatrix}
-{\boldsymbol \chi} \\
Q_{\text{tot}}
\end{bmatrix}.
\label{eq:linear_qeq}
\end{equation}
Here, ${\boldsymbol \chi}^\top = [ \chi_1, \chi_2, \ldots, \chi_N]$, ${\bf 1}^\top = [1,1,\ldots, 1]$, and the elements of the \(N \times N\) Coulomb interaction matrix \(\mathbf{C}\) given by \( C_{ij} = \delta_{ij} u_i + (\delta_{ij} - 1)\phi(r_{ij})\), where \(\delta_{ij}\) is the Kronecker delta operator and \(\phi(r_{ij})\) represents the electrostatic potential between particles \(i\) and \(j\) as given in \cref{eq:ewald_general}. 
The ground-state charges, \(\mathbf{q}^{\text{min}}\), obtained by solving \cref{eq:linear_qeq}, then determine the Born-Oppenheimer potential in \cref{eq:BO_Min}.


Calculating ground state charges by solving \cref{eq:linear_qeq}, which is especially needed for large molecular systems with periodic boundary conditions, typically requires the use of an iterative linear solver such as the generalized minimal residual (GMRES) algorithm \cite{YSaad_86}, the conjugate gradient (CG) method \cite{MRHestenes52,hestenes,CPaige75,ANakano97} or related approaches \cite{ANiklasson20,ANiklasson21}. While iterative solvers aid scalability, the solver demands tight convergence for stable simulation with sufficiently accurate and conservative forces, leading to numerous iterations and a significant computational overhead.

Each iteration involves calculating a Coulomb-like potential, for example, using an Ewald summation algorithm. This is repeated until a sufficiently converged solution is found. Convergence can be accelerated by employing a preconditioner \cite{o2019performance}, but constructing the preconditioner matrix is computationally expensive, particularly when the Coulomb potential does not utilize tapering to sparsify \(\mathbf{C}\).

In the absence of tapering and sparsity, the matrix \(\mathbf{C}\) in \cref{eq:linear_qeq} can be defined implicitly by its action on a vector \(\mathbf{v}\) instead of constructing the matrix \(\mathbf{C}\) explicitly, which would require \(O(N^2)\) in memory and operations. The operation  \(\mathbf{Cv}\) corresponds to the  Coulomb potential from some charges in the vector \(\mathbf{v}\), which can be calculated using the Ewald summation algorithm. In an iterative solver, this operation appears for a new vector \(\mathbf{v}\) in each iteration without materializing the full matrix \(\mathbf{C}\) explicitly, enabling efficient memory usage and improved performance. Instead of explicitly constructing \(\mathbf{C}\), the Ewald summation algorithms can be used to compute the matrix-vector product \(\mathbf{Cv}\) directly. The action of $\mathbf{C}$ can be expressed as the derivative of the electrostatic energy, that is, using
\[
\mathbf{C} \mathbf{v} = {\bf u} \circ\mathbf{v} + \dfrac{\partial E_{\text{Coul}}}{\partial \mathbf{v}},
\]
where $E_{\text{Coul}}$ is the Coulomb energy calculated for the charge vector ${\mathbf{v}}$ and corresponding chemical hardness vector ${\mathbf{u}}$. Here, \( \circ \) denotes the Hadamard (element-wise) product, meaning that each component of the resulting vector is given by $(\mathbf{u} \circ \mathbf{v})_i = u_i v_i$.
By utilizing our fast implementations of regular Ewald sum or the PME method, the computational and memory costs of the iterative solver can then be reduced through a matrix-free approach to the matrix-vector operation.

For a sufficiently tightly converged solution, the Born-Oppenheimer potential, $U_{\rm BO}({\bf R})$, given by the constrained minimization in \cref{eq:BO_Min}, determines the dynamics through Newton's equation of motion,
\begin{equation}
    m_i {\bf \ddot r}_i = - \nabla_i U_{\rm BO}({\bf R}).
\end{equation}
The molecular trajectories can then be generated from a step by step integration of Newton's equation, using for example the leapfrog velocity Verlet scheme.

\subsection{Shadow Born-Oppenheimer MD for a Charge Equilibration Model}
Another approach to accelerate the charge equilibration is to reformulate the underlying problem. Rather than computing approximate ground-state charges and forces through iterative optimization of a constrained energy function corresponding to an {\it exact} regular Born-Oppenheimer potential, we take an alternative approach. We construct an approximate {\it shadow} energy function for which an exact constrained minimization can be performed directly, without iterations, while remaining valid for an underlying approximate {\it shadow} Born-Oppenheimer potential. This shadow MD approach not only offers significant computational speedup but also enhances stability and long-term energy conservation, as the forces are the exact conservative forces consistent with the underlying shadow potential. 
The shadow MD concept is explored within the XL-BOMD framework \cite{niklasson2021extended,Niklasson23}, which is closely related to backward error analysis or the shadow Hamiltonian approach \cite{HYoshida90,CGrebogi90,SToxvaerd94,GJason00,ShadowHamiltonian,SToxvaerd12,KDHammonds20} used to analyze and design new integration schemes for classical Hamiltonian dynamics. 
Originally XL-BOMD was developed for self-consistent quantum-mechanical molecular dynamics simulations \cite{ANiklasson08,niklasson2021extended}. Only recently was the XL-BOMD formalism reformulated and adapted to flexible QEQ models for shadow MD simulations \cite{ANiklasson21,ANiklasson21b,JGoff23}.

In our shadow Born-Oppenheimer molecular dynamics for the charge equilibration model we replace the charge-dependent energy function in \cref{eq:qeq_energy} by an approximate shadow energy function,
\begin{equation}
    {\cal E}({\bf R,q,n}) =  \sum_i q_i \chi_i + \frac{1}{2} \sum_i q_i^2 u_i + \frac{1}{2} \sum_{i=1}^{N} \sum_{j \neq i} (2q_i-n_i)  \phi(\mathbf{r}_{ij})n_j, \label{eq:qeq_energy}
\end{equation}
which is given by a partial linearization of ${E}({\bf R,q})$ around an approximate ground state solution, ${\bf n} \approx {\bf q}^{\rm min}$. Our shadow Born-Oppenheimer potential is the given by the constrained minimization of this approximate energy function, i.e.
\begin{equation}
    {\cal U}_{\rm BO}({\bf R},{\bf n}) = V_{\rm short}({\bf R}) + \min_{\bf q} \Big\{{\cal E}({\bf R,q,n}) \Big \vert \sum_i q_i = Q_{\rm tot} \Big\}.
    \label{eq:SBO_Min}
\end{equation}
The advantage with this approximate shadow Born-Oppenheimer potential is that the constrained minimization (or rather finding a stationary solution) is straightforward to solve exactly without any iterative procedure. The ${\bf n}$-dependent relaxed ground state charge, ${\bf q}[{\bf n}]$, is given from the solution of the quasi-diagonal system of equations, 
\begin{equation}
\begin{bmatrix}
{u_1}& 0 & \dots & 0 &  1\\
0 & {u_1}&  0 & \vdots & 1\\
\vdots & 0 &  \ddots & \vdots & \vdots\\
0 & \ldots &  0 & u_N & 1\\
1 & 1 &  \ldots & 1 & 0\\
\end{bmatrix}
\begin{bmatrix}
q_1[{\bf n}] \\
q_2[{\bf n}]\\
\vdots\\
q_N[{\bf n}]\\
\lambda
\end{bmatrix}
=
\begin{bmatrix}
-{ \chi_1} - \sum_j^{j \ne 1} \phi({\bf r}_{ij}) n_j\\
-{ \chi_2} - \sum_j^{j \ne 1} \phi({\bf r}_{ij}) n_j\\
\ldots\\
-{ \chi_N} - \sum_j^{j \ne N} \phi({\bf r}_{ij}) n_j\\
Q_{\text{tot}}
\end{bmatrix},
\label{eq:diag_linear_qeq}
\end{equation}
where $\lambda$ is a Lagrange multiplier that enforces the net charge constraint.
The main cost solving this system of equations is the summation for the $v_i^{\rm Coul} = \sum_j^{j \ne i} \phi({\bf r})_{ij}n_j$ terms, which corresponds to the Coulomb potential, $v_i^{\rm Coul}$, calculated for the charges ${\bf n}$, where  $v_i^{\rm Coul} = \partial E_{\rm Coul}[{\bf n}] \big / \partial n_i $. By taking advantage of our fast implementation of the regular Ewald summation or the PME methods, this cost is significantly reduced. In contrast to the iterative methods required for the solution of the regular Born-Oppenheimer optimization problem in \cref{eq:linear_qeq} we only need to do a single construction of the Coulomb potential to get the exact ground state solution, ${\bf q}[{\bf n}]$, to determine the shadow Born-Oppenheimer potential, ${\cal U}({\bf R,n})$. 

The accuracy of the shadow potential, compared to the exact regular Born-Oppenheimer potential, will depend on how close the approximate ground state solution, ${\bf n}$, is to the exact regular ground state, ${\bf q}^{\rm min}$. Even if we would have a good initial guess of ${\bf n}$, the accuracy would get worse as the atoms are moving during a MD simulation. To avoid this problem we can update ${\bf n}$ as the atoms are moving by propagating ${\bf n}$ as an extended dynamical field variable together with the atomic positions and velocities. We can do this through an extended Lagrangian formulation \cite{ANiklasson08,ANiklasson21b},  where the dynamical field variable, ${\bf n}(t)$, is propagated through an harmonic oscillator that is centered around the ground state solution, ${\bf q}^{\rm min}$, or rather our best available approximate solution, ${\bf q}[{\bf n}]$, from \cref{eq:diag_linear_qeq}. This is in the spirit of Car-Parrinello molecular dynamics \cite{RCar85,MSprik88}, but the Lagrangian dynamics is different.

The extended Lagrangian is defined by
\begin{align}
{\cal L}({\bf R},{\bf \dot R}, {\bf n}, {\bf \dot n}) &= \frac{1}{2} \sum_i m_i \vert{\bf \dot r}_i\vert^2 - {\cal U}_{\rm BO} ({\bf R,n}) + \frac{1}{2}\mu \sum_i {\dot n}_i^2 \notag \\
&\quad - \frac{1}{2} \mu \omega^2 \sum_{ij} \left(q_i[{\bf n}] - n_i \right)T_{ij}\left(q_j[{\bf n}] - n_j \right),
\end{align}
where the first two terms are the kinetic and shadow potential energy and the last two terms are the kinetic and potential energies of the harmonic oscillator determining the dynamics of the extended electronic degrees of freedom. 
Here $m_i$ is the atomic mass of atom $i$ at position, ${\bf r}_i$, $\mu$ is the fictitious mass of partial charge $n_i$ with time derivative, ${\dot n}_i$. The frequency of the harmonic oscillator is $\omega$, and ${\bf T} = \{T_{ij}\} = {\bf K}^\top {\bf K}$ is a symmetric metric tensor of the harmonic well, which we define as the square of a kernel, ${\bf K} = {\bf J}^{-1}$, given by the inverse of the Jacobian, ${\bf J}$, of the residual function, $({\bf q}[{\bf n}] - {\bf n})$, i.e.\ where
\begin{equation}
    J_{ij} = \frac{\partial \left(q_i[{\bf n}] - n_i\right)}{\partial n_j}.
\end{equation}
The metric tensor is chosen to keep ${\bf n}(t)$ moving as close as possible around the exact regular Born-Oppenheimer ground state, ${\bf q}^{\rm min}$.

The Euler-Lagrange's equations of motion of the extended Lagragnian can then be derived in a classical adiabatic limit, where we assume the extended dynamical charge degrees of freedom is fast and light compared to the slower moving and heavier nuclear degrees of freedom, i.e. in analogue to the original Born-Oppenheimer approximation. We include this adiabatic assumption by letting $\omega \rightarrow \infty$, and $\mu \rightarrow 0$, while $\mu \omega = {\rm constant}$. In this adiabatic mass-zero limit we get the two coupled equations of motion,
\begin{equation}
    m_i {\bf \ddot r}_i = - \nabla_i {\cal U}({\bf R,n})\big \vert_n,
\end{equation}
\begin{equation}
    {\bf \ddot n} = - \omega^2 {\bf K}\left({\bf q}[{\bf n}] - {\bf n} \right).
\end{equation}
The first equation is similar to regular Born-Oppenheimer MD, but the cost of iteratively finding the sufficiently converged ground state charges to determine the potential and the forces is avoided. This procedure is instead replaced by the calculation of the shadow potential and forces, which are determined by the fully relaxed charges, ${\bf q}[{\bf n}]$, with a direct quasi-analytical solution, requiring only a single Coulomb potential calculation in each MD integration time step. The second equation we may rewrite as 
\begin{equation}
    {\bf J}{\bf \ddot n} = - \omega^2 \left({\bf q}[{\bf n}] - {\bf n} \right), \label{eq:Jnddot}
\end{equation}
which can be solved with a Krylov subspace approximation, e.g.\ the GMRES algorithm \cite{YSaad_86}. In this case ${\bf J}$ becomes similar to a modified Coulomb matrix ${\bf C}$ acting on ${\bf \ddot n}$. This may all look like we simply have replaced the original charge relaxation problem, in \cref{eq:linear_qeq}, with an equally hard problem of finding the acceleration (or forces) for ${\bf n}(t)$. In this case not much would have been gained. However, the difference is significant. With the shadow MD approach we have a shadow potential with matching conservative, exact forces, which normally requires a very tight convergence in regular Born-Oppenheimer MD, whereas the accuracy in how we integrate the extended dynamical variable ${\bf n}(t)$ based on Eq.\ (\ref{eq:Jnddot}) is of less importance, as long as ${\bf n}(t)$ stays reasonably close to the ground state. The convergence tolerance in the solution of Eq.\ (\ref{eq:Jnddot}) can therefore be kept loose. To demonstrate the efficiency of our shadow Born-Oppenheimer MD for QEQ models we can do a direct comparison to the relative convergence tolerance (and the number of Coulomb potential calculations or Ewald summations) required for regular Born-Oppenheimer MD in the solution of the charge equilibration equation, Eq.\ (\ref{eq:linear_qeq}). This charge equilibration equation is quite similar to 
Eq.\ (\ref{eq:Jnddot}) and in both cases we can use the same iterative linear solver, e.g.\ the GMRES algorithm, which allows a direct one-to-one comparison.

\section{Validation and Performance}
To create a training dataset, a subset of water molecules was extracted from the ANI-2x dataset \cite{devereux2020extending} by filtering out unrelated molecules, resulting in a dataset of approximately 330,000 molecules consisting of H and O atoms. This dataset was split into training, test and validation sets with an 80-10-10 ratio. The QEQ parameters, \({\boldsymbol \chi}\) and \({\bf u}\) (constants of each atom type H or O), were then trained against the dipole moments of the molecules in the training set. Using the optimized parameters, electrostatic energies and forces are then calculated based on the equilibrated ground-sate charges, ${\bf q}^{\rm min}$ derived from the QEQ model. A machine learning potential model was subsequently trained on the differences between the target energy and forces and those computed using the QEQ charges. This delta-trained correction, $V_{\rm short}({\bf R})$, serves as the short-range charge-independent component of the Born-Oppenheimer potential enabling MD simulations to investigate various properties of the charge equilibration methods. The regular Born-Oppenheimer potential is then given as
\begin{equation}
    U_{\text{BO}}({\bf R}) = V_{\text{short}}({\bf R}) + \sum_i q_i^{\rm min} \chi_i + \frac{1}{2} \sum_i (q_i^{\rm min})^2 u_i + E_{\text{Coul}}, \label{eq:total_en_valid}
\end{equation}
where $V_{\text{short}}({\bf R}) = \sum_i v_i^{\rm short}({\bf R})$ is the short-range machine learned potential and ${\bf q}^{\rm min} \equiv \{q_i^{\rm min}\}$ are the ground state charges calculated based on \cref{eq:BO_Min}. The same parameters are used also in the shadow MD simulations with the shadow energy function and shadow Born-Oppenheimer potential in Eqs.\ (\ref{eq:qeq_energy}) and (\ref{eq:SBO_Min}). We trained the Hierarchically Interacting Particle Neural Network (HIP-NN) \cite{lubbers2018hierarchical,hippynn} architecture for the short range interactions. The details of the model architecture and training are provided in the \cref{section:model_details} of Supplementary Information (SI).

\subsection{Validation} 

\begin{table}[H]
\centering
\begin{tabular}{@{}ccll@{}}
\toprule
\multicolumn{1}{l}{Shadow MD} & Sol. Tol. & \# Ewald Sum. & Energy Std. Dev. (eV) \\ \midrule
\multirow{6}{*}{Yes}         & $10^{-1}$      & 4.0          & 0.00542               \\
                             & $10^{-2}$      & 5.0          & 0.00532               \\
                             & $10^{-3}$      & 6.0          & 0.00512               \\
                             & $10^{-4}$      & 7.6          & 0.00510               \\
                             & $10^{-5}$      & 8.6          & 0.00518               \\
                             & $10^{-6}$      & 10.0         & 0.00527               \\ \midrule
\multirow{6}{*}{No}          & $10^{-1}$      & 6.0          & \textgreater 1000.0   \\
                              & $10^{-2}$      & 6.0          & 1.08831               \\
                              & $10^{-3}$      & 7.0          & 0.01550               \\
                              & $10^{-4}$      & 8.0          & 0.00771               \\
                              & $10^{-5}$      & 10.1         & 0.00520               \\
                              & $10^{-6}$      & 11.5         & 0.00506               \\ \bottomrule
\end{tabular}
\caption{Comparison of regular and shadow MD-based QEQ methods across different solver tolerances. Simulations of 100 water molecules are conducted in the microcanonical NVE ensemble for 100 ps with a 0.4 fs time step. The number of Ewald summations per time step is given, averaged over the simulation. Energy divergence is quantified as the standard deviation of the total energy (kinetic + potential) over the duration of the simulation.}
\label{table:solver_valid}
\end{table}

\begin{figure}
\centering
        \includegraphics[width=0.70\textwidth]{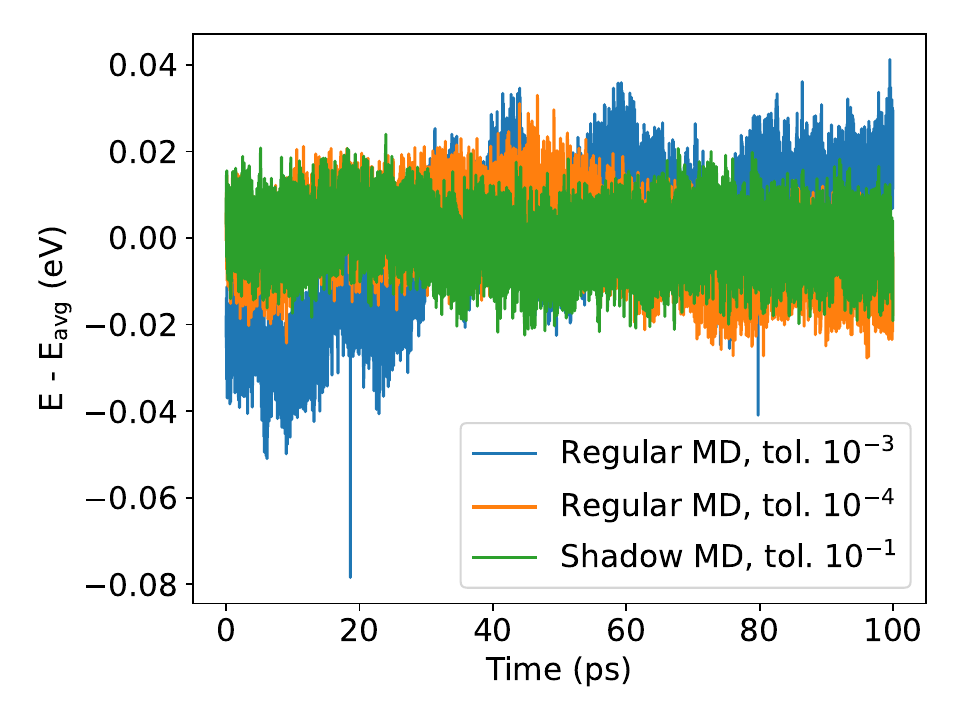}
    \caption{Example NVE simulations from \cref{table:solver_valid}. Y-axis shows the deviation of total energy from the average energy ($E_{avg}$).}
    \label{fig:nve_total_energy}
\end{figure}

To validate that the total energy conserved for the shadow and regular MD simulations with different solver tolerances, we run various microcanonical NVE simulations of 100 water molecules under periodic boundary conditions with a time step of 0.4 fs. For all of the simulations, the GMRES solver is used. Jacobi preconditioner is used for regular MD runs as the diagonal entries of the matrix are not equal to one in that case. The average temperature during these simulations are around 300 Kelvin and all of the calculations are done in double precision. As seen in \cref{table:solver_valid}, while the total energy is conserved for all tolerance levels for the shadow MD runs, the stability of regular MD runs decreases after 0.001 solver tolerance. Here, the solver tolerance refers to the maximum allowed relative residual norm, \(\| b - Ax \| / \| b \|\), for convergence. There is also a drastic difference in the number of Ewald summations. A comparison of various MD simulations are shown in Fig.\ \ref{fig:nve_total_energy}.

\begin{figure}[H]
    \centering
    \begin{subfigure}[b]{0.49\textwidth}
        \includegraphics[width=\textwidth]{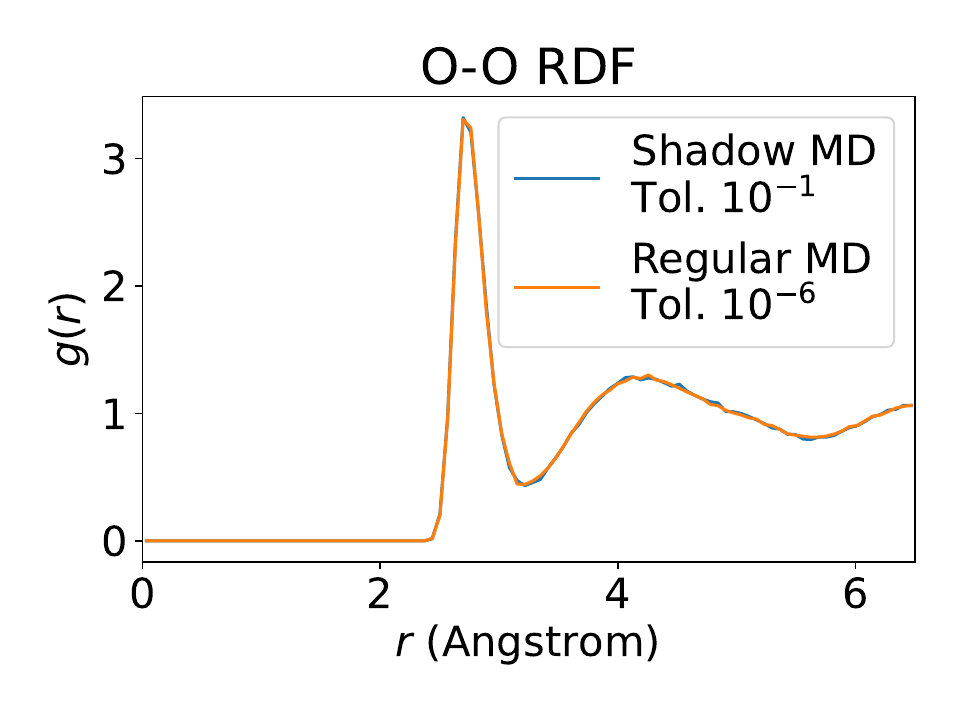}
    \end{subfigure}
    \hfill
    \begin{subfigure}[b]{0.49\textwidth}
        \includegraphics[width=\textwidth]{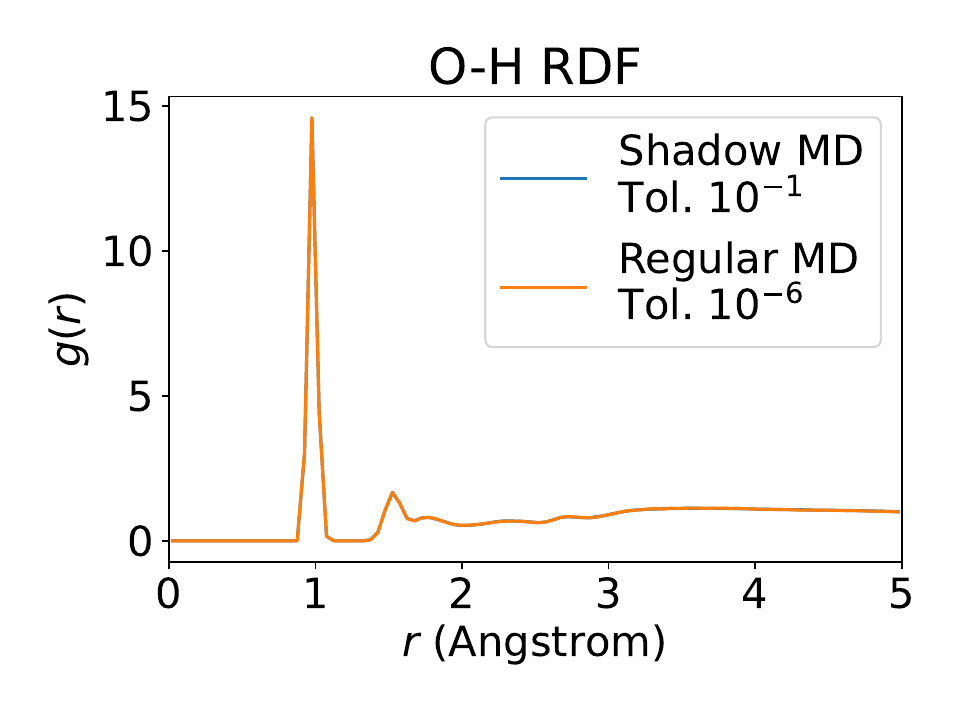}
    \end{subfigure}

    \caption{Radial distribution functions for O-O (left) and O-H (right) pairs.}
    \label{fig:RDF_comparison}
\end{figure}

\begin{figure}[H]
    \centering
    \begin{subfigure}[b]{0.49\textwidth}
        \includegraphics[width=\textwidth]{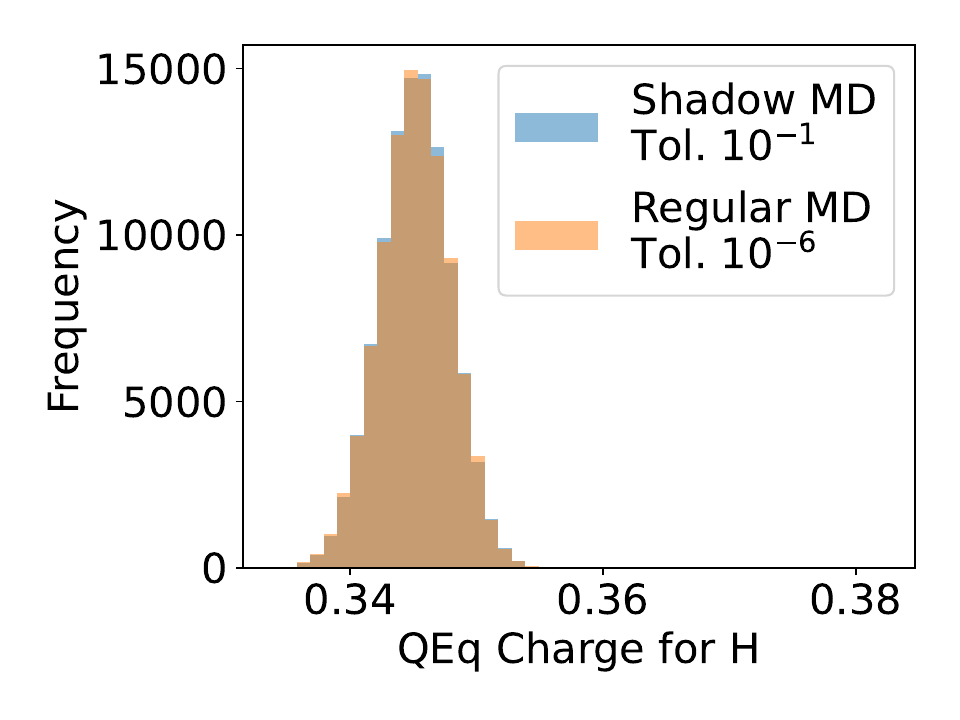}
    \end{subfigure}
    \hfill
    \begin{subfigure}[b]{0.49\textwidth}
        \includegraphics[width=\textwidth]{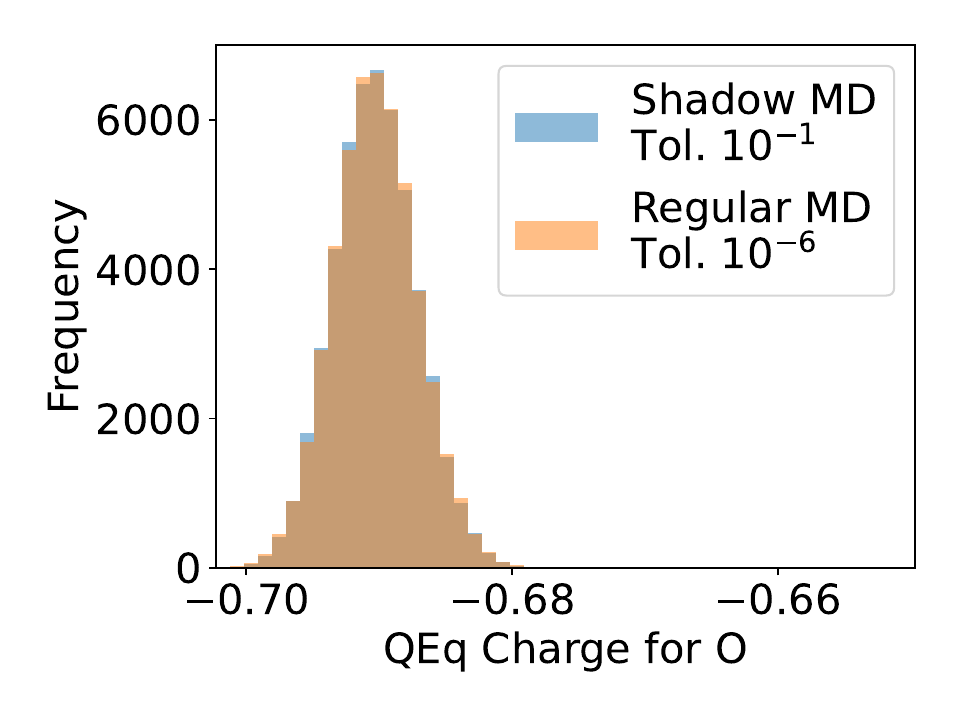}
    \end{subfigure}

    \caption{Charge distribution of hydrogen (left) and oxygen (right) atoms.}
    \label{fig:charge_comparison}
\end{figure}

In addition to energy conservation, we also compared the radial distribution functions (RDFs) for O-O and O-H pairs (\cref{fig:RDF_comparison}) as well as the charge distributions of H and O atoms (\cref{fig:charge_comparison}) across the MD runs mentioned earlier. As shown in the corresponding figures, no significant differences could be observed in these properties between shadow MD and regular MD simulations.

\subsection{Performance}
\label{subsec:performance}

In this section, we examine the scalability and efficiency of our high-performance implementations of the components required for MD simulations with charge equilibration. We focus on evaluating the computational performance of the key components essential for charge equilibration: the neighbor list generation, the traditional Ewald summation, and the PME method. Unless stated otherwise, the Ewald calculations use real space cutoff of 10 \AA\ and Ewald accuracy of $5.0 \times 10^{-4}$. All calculations except the neighbor list construction are performed in double precision.

\subsubsection{Neighbor List}

\begin{figure}[H]
    \centering
    \begin{subfigure}[b]{0.49\textwidth}
        \includegraphics[width=\textwidth]{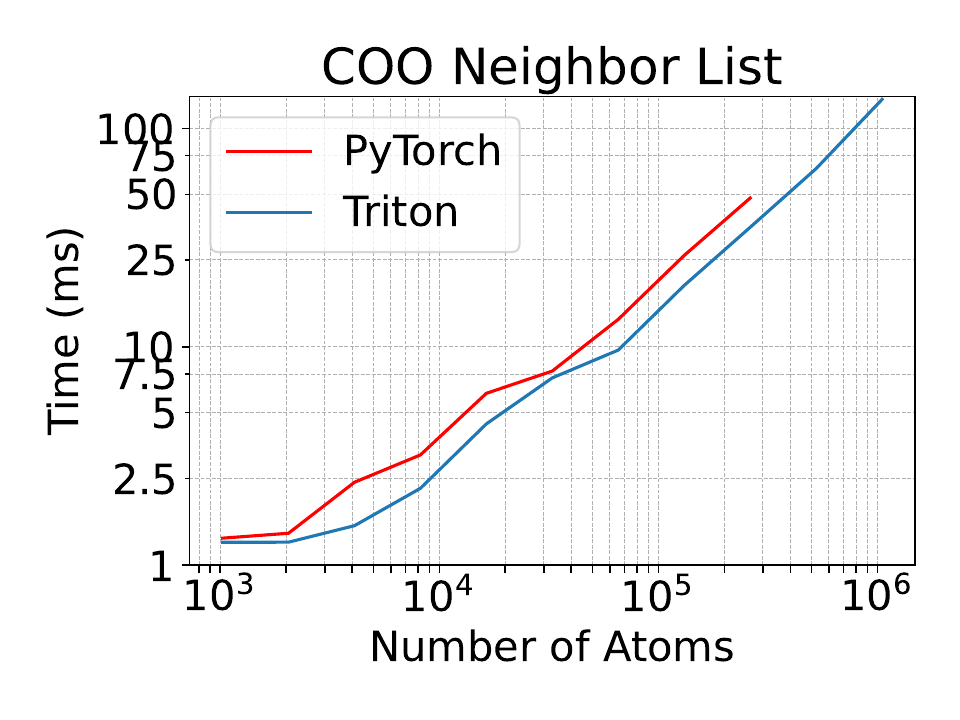}
    \end{subfigure}
    \hfill
    \begin{subfigure}[b]{0.49\textwidth}
        \includegraphics[width=\textwidth]{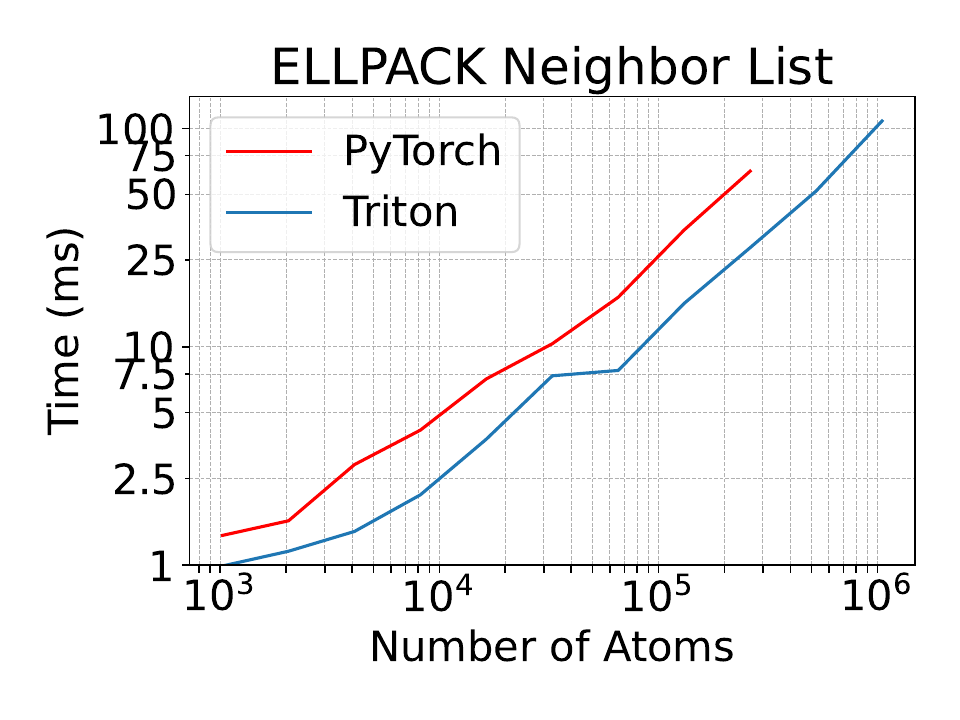}
    \end{subfigure}

    \caption{Performance of neighbor list generation for the COO format (left panel) and the ELLPACK format (right panel) on an A100 GPU with 40 GB VRAM using randomly generated water-like systems. Missing data points indicate GPU memory limitations.}
    \label{fig:neighbor_list_perf}
\end{figure}

As shown in \cref{fig:neighbor_list_perf}, our implementation efficiently generates neighbor lists in both the ELLPACK (right panel) and COO formats (left panel) on a GPU. However, the pure PyTorch implementation encounters memory constraints due to extensive vectorization. This limitation is resolved by offloading the cell-based neighbor search logic to Triton, which leverages shared memory more efficiently and allows explicit looping over neighboring cells for the given atoms. This adjustment results in up to a 2× speedup for the ELLPACK format and a 1.5× speedup for the COO format. Notably, only the search logic is implemented in Triton, while the remainder of the code remains in PyTorch. Our initial analysis suggests that this approach achieves a favorable balance between performance, development effort, and maintainability. The calculations for \cref{fig:neighbor_list_perf} are performed in single precision, as this step focuses solely on determining discrete neighbor lists and does not require high numerical precision.

\subsubsection{Ewald Summation}
\begin{figure}[H]
\centering
        \includegraphics[width=0.6\textwidth]{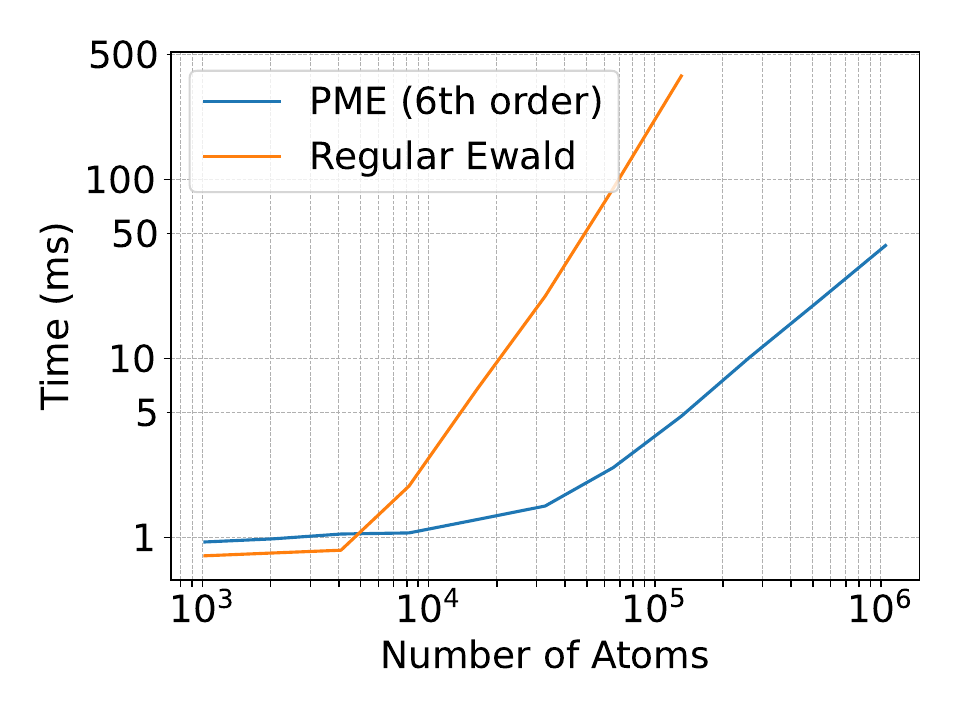}
    \caption{Comparison of regular Ewald and PME methods to calculate charge derivatives for various randomly generated water-like systems. The calculations are done in double precision on an A100 GPU with 40 GB VRAM.}
    \label{fig:ewald_performance}
\end{figure}

As previously mentioned, we provide implementations for both regular Ewald and PME methods, which can be used for charge equilibration and force calculations. Both approaches are tunable in terms of computational cost and accuracy. To determine the number of k-space vectors for regular Ewald summation and the number of grid points for PME, we use the same heuristics as OpenMM \cite{eastman2023openmm}. In \cref{fig:ewald_performance} we show a comparison between PME and regular Ewald summations. All runs are performed with the same target Ewald accuracy, and the PME implementation employs a 6th-order B-spline interpolation. These benchmarks show that regular Ewald summation is faster for systems with fewer than approximately 5000 atoms, whereas PME becomes significantly more efficient for larger systems. However, the exact trade-off point is system-dependent; the benchmarks presented here are based on randomly generated water-like systems. It is also worth noting that the regular Ewald summation utilizes custom Triton kernels, whereas the PME implementation is currently implemented purely in PyTorch due to the shared memory indexing  limitations in Triton.

\subsubsection{MD Simulations}
\begin{figure}[H]
    \centering
    \begin{subfigure}{0.49\textwidth}
        \centering
        \includegraphics[width=\linewidth]{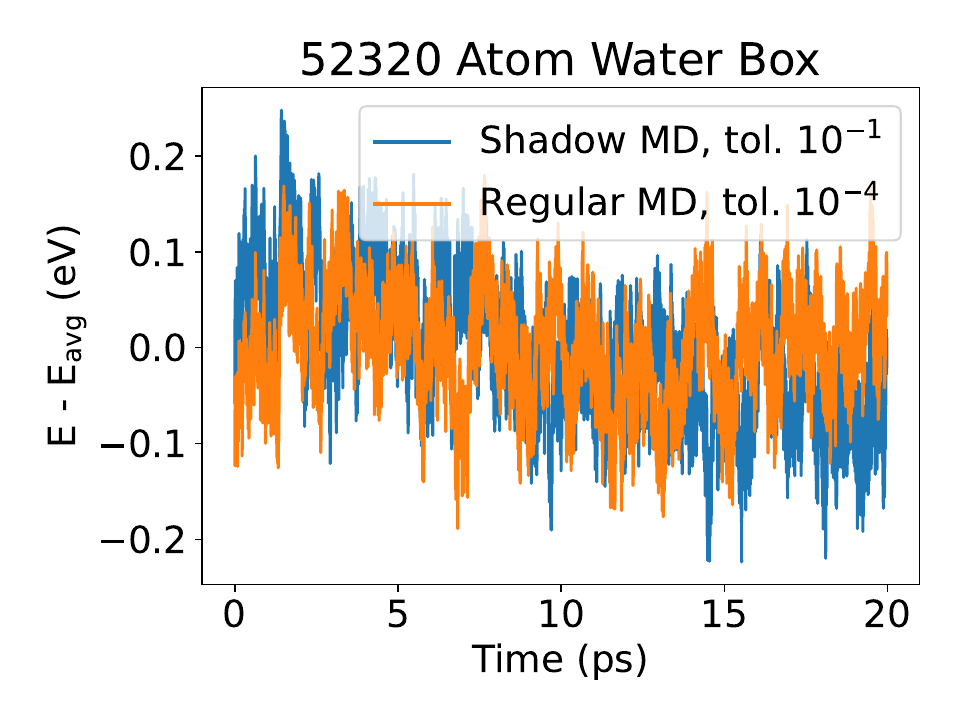}
        \label{fig:52k_nve}
    \end{subfigure}
    \hfill
    \begin{subfigure}{0.49\textwidth}
        \centering
        \includegraphics[width=\linewidth]{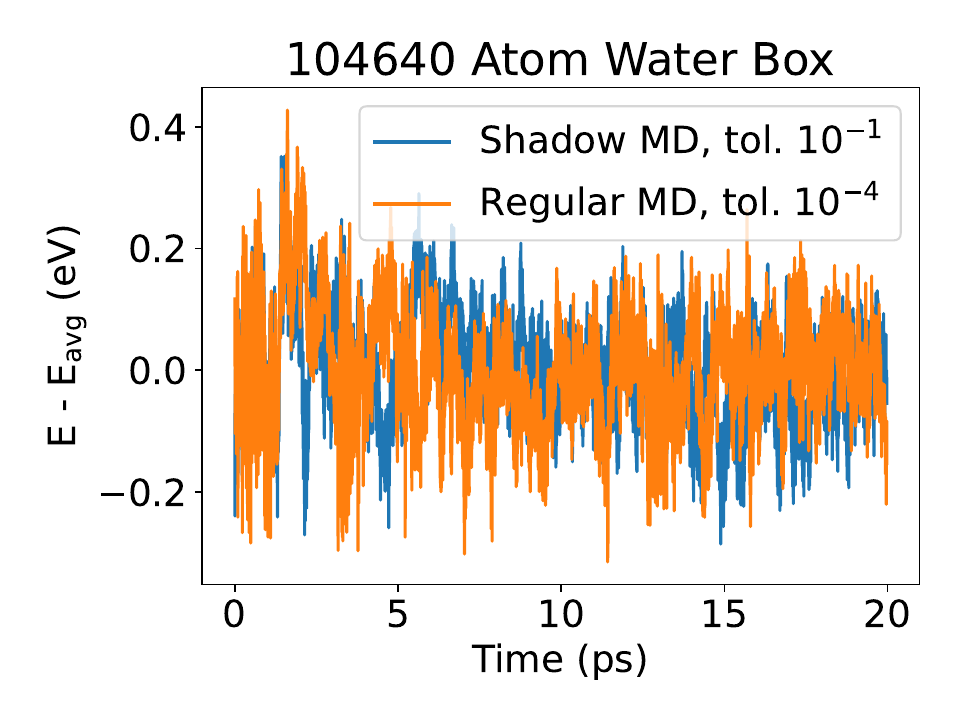}
        \label{fig:104_nve}
    \end{subfigure}

    \caption{Total energy from NVE simulations with a time step of 0.2 fs for 52,320 and 104,640 atom water boxes with dynamic charges.}
    \label{fig:large_scale_nve}
\end{figure}

To have a more realistic benchmark, we also run MD simulations using water clusters with varying number of atoms. As shown in \cref{fig:large_scale_nve}, the shadow MD approach enables relaxing the solver convergence criteria while maintaining stability. Hence, we can use 0.1 in GMRES solver tolerance for integration of the extended dynamical charges in shadow MD.  In contrast, regular MD requires a much tighter convergence to maintain stability. 

\begin{figure}[H]
\centering
        \includegraphics[width=0.95\textwidth]{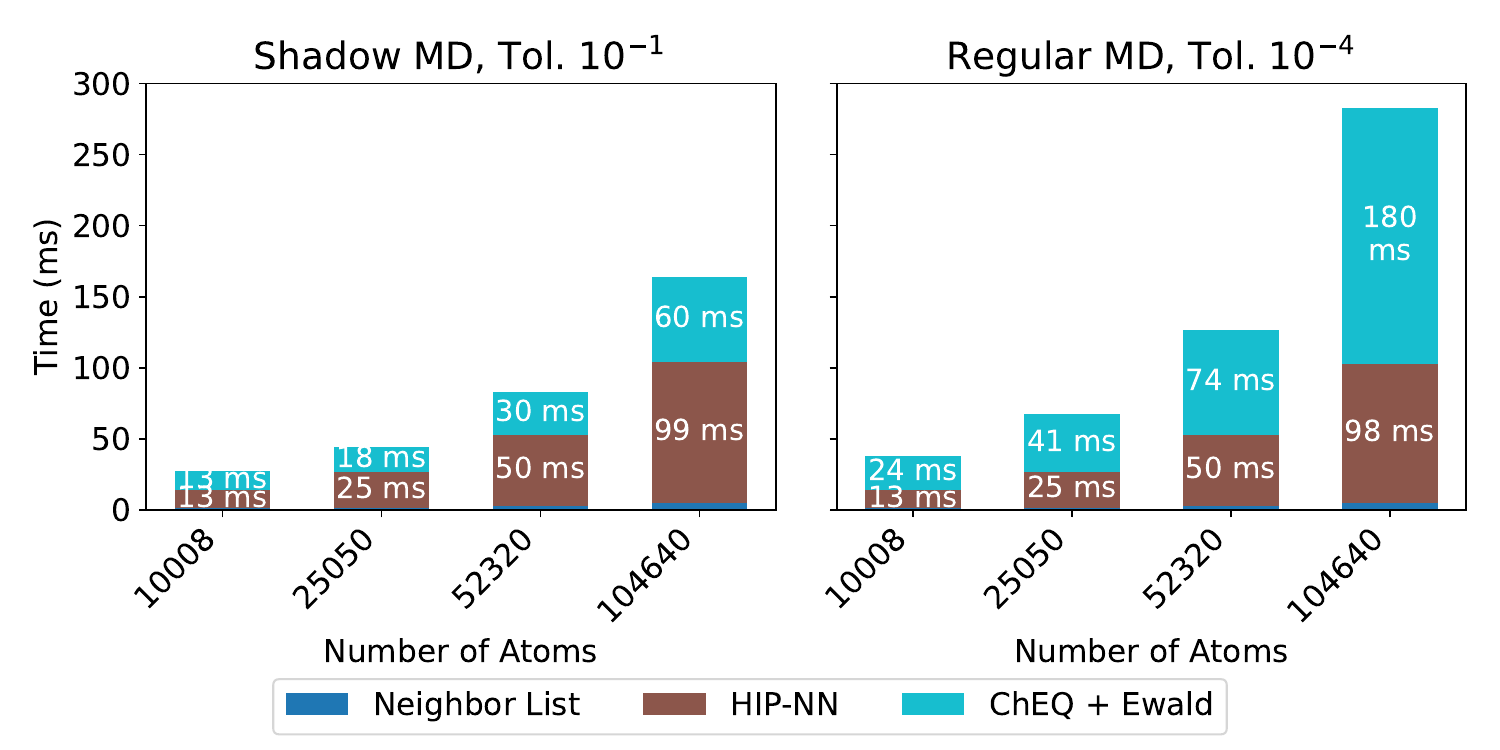}
    \caption{Average time spent on various components during MD simulations using double precision on an A100 GPU with 40 GB VRAM. The simulated systems consist of water clusters with varying numbers of atoms. Charge equilibration is performed using PME, and reneighboring occurs only when any atom moves more than half of the skin length (1.0 \AA).}
    \label{fig:MD_simulation}
\end{figure}

For large-scale MD simulations, the difference in solver convergence leads to up to $3\times$ speedup for the charge equilibration when we move from regular MD to shadow MD while they perform similar in terms of energy conservation. Due to numerical issues for the charge-independent machine-learned force field, we used a 0.2 fs time step for simulations with large water clusters. The timings of these large-scale simulations are shown in Fig.\ \ref{fig:MD_simulation}. We find that for regular MD simulations the computational cost is dominated by the PME solver, whereas the cost for the shadow MD simulations is instead governed by the charge-independent force field calculations using the HIP-NN neural network, which itself is highly optimized, taking approximately $1 \textrm{\textmu}s$ per atom per step.

\section{Conclusion}
In this work, we have introduced a novel and scalable implementation of GPU-accelerated charge equilibration for shadow molecular dynamics in Python, addressing several key limitations in the field of molecular simulations. Our approach leverages advanced computational frameworks, including PyTorch and Triton, to deliver high-performance implementations of essential components such as neighbor list generation, regular Ewald summation and PME methods.

Our shadow MD approach significantly reduces the computational cost by relaxing solver convergence requirements, maintaining energy conservation and physical accuracy in MD simulations while achieving up to a \(3\times\) speedup over traditional methods. Validation of physical observables such as radial distribution functions and charge distributions confirmed the accuracy of our implementation.

Beyond performance improvements, our framework prioritizes ease of use, maintainability, and extendability. The Python-based design ensures seamless integration with machine-learning potentials, making it highly accessible for researchers and developers. The modular architecture enables straightforward modifications and extensions, allowing users to incorporate new models, alternative electrostatics schemes, and additional optimizations with minimal effort.

Future work will focus on extending this framework to support multi-GPU molecular dynamics for larger-scale simulations, incorporating multipole expansions to enhance the accuracy of electrostatic models, and exploring alternative damping functions to improve charge equilibration efficiency. Our methods are publicly available as open-source software at \href{https://github.com/LANL/sedacs}{https://github.com/LANL/sedacs}, which facilitates their integration into commonly used molecular-dynamics codes; future work also will emphasize this integration. These advancements will further broaden the applicability of our approach to more complex and heterogeneous systems.

\section{Acknowledgements}

This work is supported by the U.S. Department of Energy Office of Basic Energy Sciences (FWP LANLE8AN), the Los Alamos National Laboratory LDRD program and by the U.S. Department of Energy through the Los Alamos National Laboratory Los Alamos National Laboratory is operated by Triad National Security, LLC, for the National Nuclear Security Administration of the U.S. Department of Energy Contract No. 892333218NCA000001.

\bibliography{References}

\begin{thebibliography}{1}

\bibitem{diederik2014adam}
P~Kingma Diederik.
\newblock Adam: A method for stochastic optimization.
\newblock {\em (No Title)}, 2014.

\bibitem{SIlubbers2018hierarchical}
Nicholas Lubbers, Justin~S Smith, and Kipton Barros.
\newblock Hierarchical modeling of molecular energies using a deep neural network.
\newblock {\em The Journal of chemical physics}, 148(24), 2018.

\end{thebibliography}


\begin{thebibliography}{10}

\bibitem{hohen}
P.~Hohenberg and W.~Kohn.
\newblock Inhomgenous electron gas.
\newblock {\em Phys. Rev.}, 136:B:864--B871, 1964.

\bibitem{KohnSham65}
W~Kohn and L.~J. Sham.
\newblock Self-consistent equations including exchange and correlation effects.
\newblock {\em Phys. Rev.}, 140(4A):1133, 1965.

\bibitem{RParr89}
R.~G. Parr and W.~Yang.
\newblock {\em Density-functional theory of atoms and molecules}.
\newblock Oxford University Press, Oxford, 1989.

\bibitem{ROJones89}
R.~O. Jones and O.~Gunnarsson.
\newblock The density functional formalism, its applications and prospects.
\newblock {\em Rev. Mod. Phys.}, 61:689--746, Jul 1989.

\bibitem{RMDreizler90}
R.M. Dreizler and K.U. Gross.
\newblock {\em Density-functional theory}.
\newblock Springer Verlag, Berlin Heidelberg, 1990.

\bibitem{JPerdew92}
John~P. Perdew, J.~A. Chevary, S.~H. Vosko, Koblar~A. Jackson, Mark~R. Pederson, D.~J. Singh, and Carlos Fiolhais.
\newblock Atoms, molecules, solids, and surfaces: Applications of the generalized gradient approximation for exchange and correlation.
\newblock {\em Phys. Rev. B}, 46:6671--6687, Sep 1992.

\bibitem{Becke93}
A.~D. Becke.
\newblock A new mixing of hartree-fock and local density-functional theories.
\newblock {\em J. Chem. Phys.}, 98:1372--1377, 1993.

\bibitem{WKohn99}
W.~Kohn.
\newblock Nobel lecture: Electronic structure of matter---wave functions and density functionals.
\newblock {\em Rev. Mod. Phys.}, 71:1253--1266, Oct 1999.

\bibitem{monticelli2013force}
Luca Monticelli and D~Peter Tieleman.
\newblock Force fields for classical molecular dynamics.
\newblock {\em Biomolecular simulations: Methods and protocols}, pages 197--213, 2013.

\bibitem{senftle2016reaxff}
Thomas~P Senftle, Sungwook Hong, Md~Mahbubul Islam, Sudhir~B Kylasa, Yuanxia Zheng, Yun~Kyung Shin, Chad Junkermeier, Roman Engel-Herbert, Michael~J Janik, Hasan~Metin Aktulga, et~al.
\newblock The reaxff reactive force-field: development, applications and future directions.
\newblock {\em npj Computational Materials}, 2(1):1--14, 2016.

\bibitem{FJVesely77}
F.~J. Vesely.
\newblock N-particle dynamics of polarizable stockmayer-type molecules.
\newblock {\em J. Comput. Phys.}, 24:361--371, 1977.

\bibitem{WJMortier86}
Wilfried~J. Mortier, Swapan~K. Ghosh, and S.~Shankar.
\newblock Electronegativity-equalization method for the calculation of atomic charges in molecules.
\newblock {\em Journal of the American Chemical Society}, 108(15):4315--4320, 1986.

\bibitem{MSprik88}
Michiel Sprik and Michael~L. Klein.
\newblock A polarizable model for water using distributed charge sites.
\newblock {\em The Journal of Chemical Physics}, 89(12):7556--7560, 1988.

\bibitem{MSprik90}
M.~Sprik.
\newblock Computer simulation of the dynamics of induced polarization fluctuations in water.
\newblock {\em J. Chem. Phys.}, 95:2283--2291, 1990.

\bibitem{AKRappe91}
Anthony~K. Rappe and William A.~Goddard III.
\newblock Charge equilibration for molecular dynamics simulations.
\newblock {\em J. Phys. Chem}, 95(8):3358--3363, 1991.

\bibitem{DVanBelle92}
D.~Van~Belle, M.~Froeyen, G.~Lippens, and S.~J. Wodak.
\newblock Molecular dynamics simulation of polarizable water by an extended lagrangian method.
\newblock {\em Mol. Phys.}, 77:239--266, 1992.

\bibitem{WSRick94}
S.~W. Rick, S.~J. Stuart, and B.~J. Berne.
\newblock Dynamical fluctuating charge force fields: Application to liquid water.
\newblock {\em J. Chem. Phys.}, 101:6141--6156, 1994.

\bibitem{TAHalgren01}
T.~A. Halgren and D.~Wolfgang.
\newblock Polarizable force fields.
\newblock {\em Curr. Opin. Struct. Biol.}, 11:236--242, 2001.

\bibitem{GLamoureaux03}
G.~Lamoureux and B.~T. Roux.
\newblock Modeling induced polarization with classical drude oscillators: Theory and molecular dynamics simulation algorithm.
\newblock {\em J. Chem. Phys.}, 119:3025--3039, 2003.

\bibitem{GAKaminsky04}
G.~A. Kaminski, H.~A. Stern, B.~J. Berne, and R.~A. Friesner.
\newblock Development of an accurate and robust polarizable molecular mechanics force field from ab initio quantum chemistry.
\newblock {\em J. Phys. Chem. A}, 108:621--627, 2004.

\bibitem{PEMLopes09}
P.~E.~M. Lopes, B.~Roux, and A.~D.~Jr. MacKerell.
\newblock Molecular modeling and dynamics studies with explicit inclusion of electronic polarizability: theory and applications.
\newblock {\em Theor. Chem. Acc.}, 124:11--28, 2009.

\bibitem{PCieplak09}
P.~Cieplak, F.-Y. Dupradeau, Y.~Duan, and J.~Wang.
\newblock Polarization effects in molecular mechanical force fields.
\newblock {\em J. Phys.: Condens. Matter}, 21:333102, 2009.

\bibitem{SNaserifar17}
Saber Naserifar, Daniel~J. Brooks, William~A. Goddard, and Vaclav Cvicek.
\newblock Polarizable charge equilibration model for predicting accurate electrostatic interactions in molecules and solids.
\newblock {\em The Journal of Chemical Physics}, 146(12):124117, 2017.

\bibitem{JZhifeng19}
Zhifeng Jing, Chengwen Liu, Sara~Y. Cheng, Rui Qi, Brandon~D. Walker, Jean-Philip Piquemal, and Pengyu Ren.
\newblock Polarizable force fields for biomolecular simulations: Recent advances and applications.
\newblock {\em Annual Review of Biophysics}, 48(1):371--394, 2019.
\newblock PMID: 30916997.

\bibitem{SGoedecker15}
S.~Alireza Ghasemi, Albert Hofstetter, Santanu Saha, and Stefan Goedecker.
\newblock Interatomic potentials for ionic systems with density functional accuracy based on charge densities obtained by a neural network.
\newblock {\em Phys. Rev. B}, 92:045131, Jul 2015.

\bibitem{TWKo20}
Tsz~Wai Ko, Jonas~A. Finkler, Stefan Goedecker, and Jörg Behler.
\newblock A fourth-generation high-dimensional neural network potential with accurate electrostatics including non-local charge transfer.
\newblock {\em Nature Comm.}, 12:398, 2021.

\bibitem{TWKo23}
Tsz~Wai Ko, Jonas~A. Finkler, Stefan Goedecker, and J{\"o}rg Behler.
\newblock Accurate fourth-generation machine learning potentials by electrostatic embedding.
\newblock {\em J. Chem. Theory Comput.}, 19(12):3567--3579, 2023.
\newblock PMID: 37289440.

\bibitem{nomura2015extended}
Ken-ichi Nomura, Patrick~E Small, Rajiv~K Kalia, Aiichiro Nakano, and Priya Vashishta.
\newblock An extended-lagrangian scheme for charge equilibration in reactive molecular dynamics simulations.
\newblock {\em Computer Physics Communications}, 192:91--96, 2015.

\bibitem{paszke2019pytorch}
Adam Paszke, Sam Gross, Francisco Massa, Adam Lerer, James Bradbury, Gregory Chanan, Trevor Killeen, Zeming Lin, Natalia Gimelshein, Luca Antiga, et~al.
\newblock Pytorch: An imperative style, high-performance deep learning library.
\newblock {\em Advances in neural information processing systems}, 32, 2019.

\bibitem{tillet2019triton}
Philippe Tillet, Hsiang-Tsung Kung, and David Cox.
\newblock Triton: an intermediate language and compiler for tiled neural network computations.
\newblock In {\em Proceedings of the 3rd ACM SIGPLAN International Workshop on Machine Learning and Programming Languages}, pages 10--19, 2019.

\bibitem{ANiklasson21}
A.~M.~N. Niklasson.
\newblock Extended lagrangian born-oppenheimer molecular dynamics for orbital-free density functional theory and polarizable charge equilibration models.
\newblock {\em J. Chem. Phys.}, 154:0000, 2021.

\bibitem{JGoff23}
James Goff, Yu~Zhang, Christian Negre, Andrew Rohskopf, and Anders M.~N. Niklasson.
\newblock Shadow molecular dynamics and atomic cluster expansions for flexible charge models.
\newblock {\em Journal of Chemical Theory and Computation}, 19(13):4255--4272, 2023.
\newblock PMID: 37382528.

\bibitem{Niklasson23}
Anders M.~N. Niklasson and Christian F.~A. Negre.
\newblock {Shadow energy functionals and potentials in Born–Oppenheimer molecular dynamics}.
\newblock {\em The Journal of Chemical Physics}, 158(15):154105, 04 2023.

\bibitem{thompson2022lammps}
Aidan~P Thompson, H~Metin Aktulga, Richard Berger, Dan~S Bolintineanu, W~Michael Brown, Paul~S Crozier, Pieter~J In't~Veld, Axel Kohlmeyer, Stan~G Moore, Trung~Dac Nguyen, et~al.
\newblock Lammps-a flexible simulation tool for particle-based materials modeling at the atomic, meso, and continuum scales.
\newblock {\em Computer Physics Communications}, 271:108171, 2022.

\bibitem{aktulga2012parallel}
Hasan~Metin Aktulga, Joseph~C Fogarty, Sagar~A Pandit, and Ananth~Y Grama.
\newblock Parallel reactive molecular dynamics: Numerical methods and algorithmic techniques.
\newblock {\em parallel computing}, 38(4-5):245--259, 2012.

\bibitem{kaymak2023end}
Mehmet~Cagri Kaymak, Samuel~S Schoenholz, Ekin~D Cubuk, Kurt~A O’Hearn, Kenneth~M Merz~Jr, and Hasan~Metin Aktulga.
\newblock End-to-end differentiable reactive molecular dynamics simulations using jax.
\newblock In {\em International Conference on High Performance Computing}, pages 202--219. Springer, 2023.

\bibitem{gubler2024accelerating}
Moritz Gubler, Jonas~A Finkler, Moritz~R Sch\"afer, Jörg Behler, and Stefan Goedecker.
\newblock Accelerating fourth-generation machine learning potentials using quasi-linear scaling particle mesh charge equilibration.
\newblock {\em Journal of Chemical Theory and Computation}, 20(16):7264--7271, 2024.

\bibitem{ANiklasson21b}
A.~M.~N. Niklasson.
\newblock Extended lagrangian born–oppenheimer molecular dynamics: from density functional theory to charge relaxation models.
\newblock {\em Eur. Phys. J. B}, 94:164, 2021.

\bibitem{CHLi2025}
Cheng-Han Li, Mehmet~Cagri Kaymak, Maksim Kulichenko, Nicholas Lubbers, Benjamin Nebgen, Sergei Tretiak, Joshua Finkelstein, Daniel Tabor, and Anders Niklasson.
\newblock Shadow molecular dynamics with a machine learned flexible charge potential.
\newblock {\em ChemRxiv}, 2025.

\bibitem{KNomura15}
K.~Nomura, P.~E. Small, R.~K. Kalia, A.~Nakano, and P.~Vashista.
\newblock An extended-lagrangian scheme for charge equilibration in reactive molecular dynamics simulations.
\newblock {\em Comput. Phys. Comm.}, 192:91, 2015.

\bibitem{AAlbaugh15}
A.~Albaugh, O.~Demardash, and T.~Head-Gordon.
\newblock An efficient and stable hybrid extended lagrangian/self-consistent field scheme for solving classical mutual induction.
\newblock {\em J. Chem. Phys.}, 143:174104, 2015.

\bibitem{AAlbaugh18}
Alex Albaugh, Teresa Head-Gordon, and Anders M.~N. Niklasson.
\newblock {Higher-Order Extended Lagrangian Born-Oppenheimer Molecular Dynamics for Classical Polarizable Models}.
\newblock {\em Journal of Chemical Theory and Computation}, 14(2):499--511, 2018.
\newblock PMID: 29316388.

\bibitem{ILeven19}
I.~Leven and T.~Head-Gordon.
\newblock Inertial extended-lagrangian scheme for solving charge equilibration models.
\newblock {\em Phys. Chem. Chem. Phys.}, 21(34):18652--18659, 2019.

\bibitem{ANiklasson06}
Anders M.~N. Niklasson, C.~J. Tymczak, and M.~Challacombe.
\newblock {Time-reversible Born-Oppenheimer molecular dynamics}.
\newblock {\em Phys. Rev. Lett.}, 97:123001, 2006.

\bibitem{ANiklasson08}
Anders M.~N. Niklasson.
\newblock {Extended Born-Oppenheimer molecular dynamics}.
\newblock {\em Phys. Rev. Lett.}, 100:123004, 2008.

\bibitem{MAllen90}
M.~Allen and D.~Tildesley.
\newblock {\em Computer Simulation of Liquids}.
\newblock Oxford Science, London, 1990.

\bibitem{schoenholz2020jax}
Samuel Schoenholz and Ekin~Dogus Cubuk.
\newblock Jax md: a framework for differentiable physics.
\newblock {\em Advances in Neural Information Processing Systems}, 33:11428--11441, 2020.

\bibitem{filippone2017sparse}
Salvatore Filippone, Valeria Cardellini, Davide Barbieri, and Alessandro Fanfarillo.
\newblock Sparse matrix-vector multiplication on gpgpus.
\newblock {\em ACM Transactions on Mathematical Software (TOMS)}, 43(4):1--49, 2017.

\bibitem{toukmaji1996ewald}
Abdulnour~Y Toukmaji and John~A Board~Jr.
\newblock Ewald summation techniques in perspective: a survey.
\newblock {\em Computer physics communications}, 95(2-3):73--92, 1996.

\bibitem{JZiman79}
J.~M. Ziman.
\newblock {\em Models of Disorder: The Theoretical Physics of Homogeneously Disordered Systems}.
\newblock Cambridge University Press, Cambridge, England, 1979.

\bibitem{young2014iterative}
David~M Young.
\newblock {\em Iterative solution of large linear systems}.
\newblock Elsevier, 2014.

\bibitem{eastman2023openmm}
Peter Eastman, Raimondas Galvelis, Ra{\'u}l~P Pel{\'a}ez, Charlles~RA Abreu, Stephen~E Farr, Emilio Gallicchio, Anton Gorenko, Michael~M Henry, Frank Hu, Jing Huang, et~al.
\newblock Openmm 8: molecular dynamics simulation with machine learning potentials.
\newblock {\em The Journal of Physical Chemistry B}, 128(1):109--116, 2023.

\bibitem{darden1993particle}
Tom Darden, Darrin York, and Lee Pedersen.
\newblock Particle mesh ewald: An nlog(n) method for ewald sums in large systems.
\newblock {\em The Journal of chemical physics}, 98(12):10089--10092, 1993.

\bibitem{JCooley65}
J.~W. Cooley and J.~W. Tukey.
\newblock An algorithm for the machine calculation of complex fourier series.
\newblock {\em Math. Comput.}, 19:297--301, 1965.

\bibitem{de1978practical}
Carl De~Boor and Carl De~Boor.
\newblock {\em A practical guide to splines}, volume~27.
\newblock springer New York, 1978.

\bibitem{WHeitler27}
W.~Heitler and F.~London.
\newblock Wechselwirkung neutraler atome und homöopolare bindung nach der quantenmechanik.
\newblock {\em Z.\ Phys.}, 44:455, 1927.

\bibitem{MBorn27}
M.~Born and R.~Oppenheimer.
\newblock Zur quantenteori det molekeln.
\newblock {\em Ann.\ Phys.}, 389:475, 1927.

\bibitem{DMarx00}
D.~Marx and J.~Hutter.
\newblock {\em Modern Methods and Algorithms of Quantum Chemistry}.
\newblock ed. J. Grotendorst, John von Neumann Institute for Computing, J\"ulich, Germany, second edition, 2000.

\bibitem{MTuckerman02}
M.\~E.\ Tuckerman.
\newblock Ab initio molecular dynamics: basic concepts, current trends and novel applications.
\newblock {\em J. Phys.: Conden. Matter}, 14:1297, 2002.

\bibitem{sanderson1951interpretation}
RT~Sanderson.
\newblock An interpretation of bond lengths and a classification of bonds.
\newblock {\em Science}, 114(2973):670--672, 1951.

\bibitem{YSaad_86}
Y.~Saad and M.~H. Schultz.
\newblock Gmres: A generalized minimal residual algorithm for solving nonsymmetric linear systems.
\newblock {\em SIAM J. Sci. Stat. Comput.}, 7:856, 1986.

\bibitem{MRHestenes52}
Magnus~R. Hestene and Eduard Stiefel.
\newblock Methods of conjugate gradients for solving linear systems.
\newblock {\em Journal of Research of the National Bureau of Standards}, 49:409, 1952.

\bibitem{hestenes}
M.~R. Hestens.
\newblock {\em Conjugate direction methods in optimization}.
\newblock Springer, New York, 1980.

\bibitem{CPaige75}
C.~C. Paige and M.~A. Saunders.
\newblock Solution of sparse indefinite systems of linear equations.
\newblock {\em SIAM Journal on Numerical Analysis}, 12(4):617--629, 1975.

\bibitem{ANakano97}
Aiichiro Nakano.
\newblock Parallel multilevel preconditioned conjugate-gradient approach to variable-charge molecular dynamics.
\newblock {\em Computer Physics Communications}, 104(1):59--69, 1997.

\bibitem{ANiklasson20}
A.~M.~N. Niklasson.
\newblock Extended lagrangian born-oppenheimer molecular dynamics using a krylov subspace approximation.
\newblock {\em J. Chem. Phys.}, 152:104103, 2020.

\bibitem{o2019performance}
Kurt~A O'Hearn, Abdullah Alperen, and Hasan~Metin Aktulga.
\newblock Performance optimization of reactive molecular dynamics simulations with dynamic charge distribution models on distributed memory platforms.
\newblock In {\em Proceedings of the ACM International Conference on Supercomputing}, pages 150--159, 2019.

\bibitem{niklasson2021extended}
Anders~MN Niklasson.
\newblock Extended lagrangian born--oppenheimer molecular dynamics: from density functional theory to charge relaxation models.
\newblock {\em The European Physical Journal B}, 94(8):164, 2021.

\bibitem{HYoshida90}
H.~Yoshida.
\newblock {\em Phys. Lett. A}, 150:262, 1990.

\bibitem{CGrebogi90}
C.~Grebogi, S.~M. Hammel, J.~A. Yorke, and T.~Saur.
\newblock {\em Phys. Rev. Lett.}, 65:1527, 1990.

\bibitem{SToxvaerd94}
S.~Toxvaerd.
\newblock {\em Phys. Rev. E}, 50:2271, 1994.

\bibitem{GJason00}
Jason Gans and David Shalloway.
\newblock Shadow mass and the relationship between velocity and momentum in symplectic numerical integration.
\newblock {\em Phys. Rev. E}, 61:4587--4592, Apr 2000.

\bibitem{ShadowHamiltonian}
Stephen~D. Bond and Benedict~J. Leimkuhler.
\newblock {\em Molecular dynamics and the accuracy of numerically computed averages}.
\newblock Cambride University Press, United Kingdom, 2007.

\bibitem{SToxvaerd12}
S.~Toxvaerd, O.~J. Heilmann, and J.~C. Dyre.
\newblock {\em J. Chem. Phys.}, 136:224106, 2012.

\bibitem{KDHammonds20}
K.~D. Hammonds and D.~M. Heyes.
\newblock {\em J. Chem. Phys.}, 152:024114, 2020.

\bibitem{RCar85}
R.~Car and M~Parrinello.
\newblock Unified approach for molecular dynamics and density-functional theory.
\newblock {\em Phys. Rev. Lett.}, 55:2471, 1985.

\bibitem{devereux2020extending}
Christian Devereux, Justin~S Smith, Kate~K Huddleston, Kipton Barros, Roman Zubatyuk, Olexandr Isayev, and Adrian~E Roitberg.
\newblock Extending the applicability of the ani deep learning molecular potential to sulfur and halogens.
\newblock {\em Journal of Chemical Theory and Computation}, 16(7):4192--4202, 2020.

\bibitem{lubbers2018hierarchical}
Nicholas Lubbers, Justin~S Smith, and Kipton Barros.
\newblock Hierarchical modeling of molecular energies using a deep neural network.
\newblock {\em The Journal of chemical physics}, 148(24), 2018.

\bibitem{hippynn}
The hippynn python package, 2019.

\end{thebibliography}
\bibliographystyle{unsrt}

\clearpage
\section*{Supplementary Information}
\appendix
\setcounter{figure}{0}
\setcounter{table}{0}
\renewcommand{\thesection}{S\arabic{section}}
\renewcommand{\thefigure}{S\arabic{figure}}
\renewcommand{\thetable}{S\arabic{table}}

\section{Model Training Details}
\label{section:model_details}

\begin{table}[H]
\centering
\begin{tabular}{|l|l|}
\hline
Number of Features      & 40    \\ \hline
Number of Sensitivities & 10    \\ \hline
Number of Layers        & 3     \\ \hline
Number of Interactions  & 2     \\ \hline
Tensor Sensitivity      & 0     \\ \hline
Cutoff                  & 5.0 Å \\ \hline
Min. Soft Cutoff        & 0.3 Å \\ \hline
Max. Soft Cutoff        & 3.0 Å \\ \hline
\end{tabular}
\label{table:SI_HPNN}
\caption{Hyperparameters for the HIP-NN model.}
\end{table}

The QEQ model parameters (electronegativity and chemical hardness) were trained using a batch size of 256, an initial learning rate of 0.0005, and the ADAM optimizer \citeSI{diederik2014adam}, without incorporating any learning rate scheduling.

For HIP-NN, training was conducted with a batch size of 256 and an initial learning rate of 0.0005, optimized using the ADAM optimizer. A learning rate reduction strategy was employed, halving the rate whenever the validation loss failed to improve for 5 consecutive epochs. The loss function was defined as:

\begin{equation}
\mathcal{L} = \mathcal{L}_\text{E} + \mathcal{L}_\text{F} +  \mathcal{L}_\text{H},
\end{equation}
where $\mathcal{L}_\text{E}$ represents the root mean square (RMS) of the energy error, $\mathcal{L}_\text{F}$ denotes the RMS of the force error, and $\mathcal{L}_\text{H}$ is the hierarchical loss, as defined in \citeSI{SIlubbers2018hierarchical}.

\section{Performance Comparison of Ewald Summation Methods}
\label{section:perf_details}
\begin{figure}[H]
    \centering
\includegraphics[width=\textwidth]{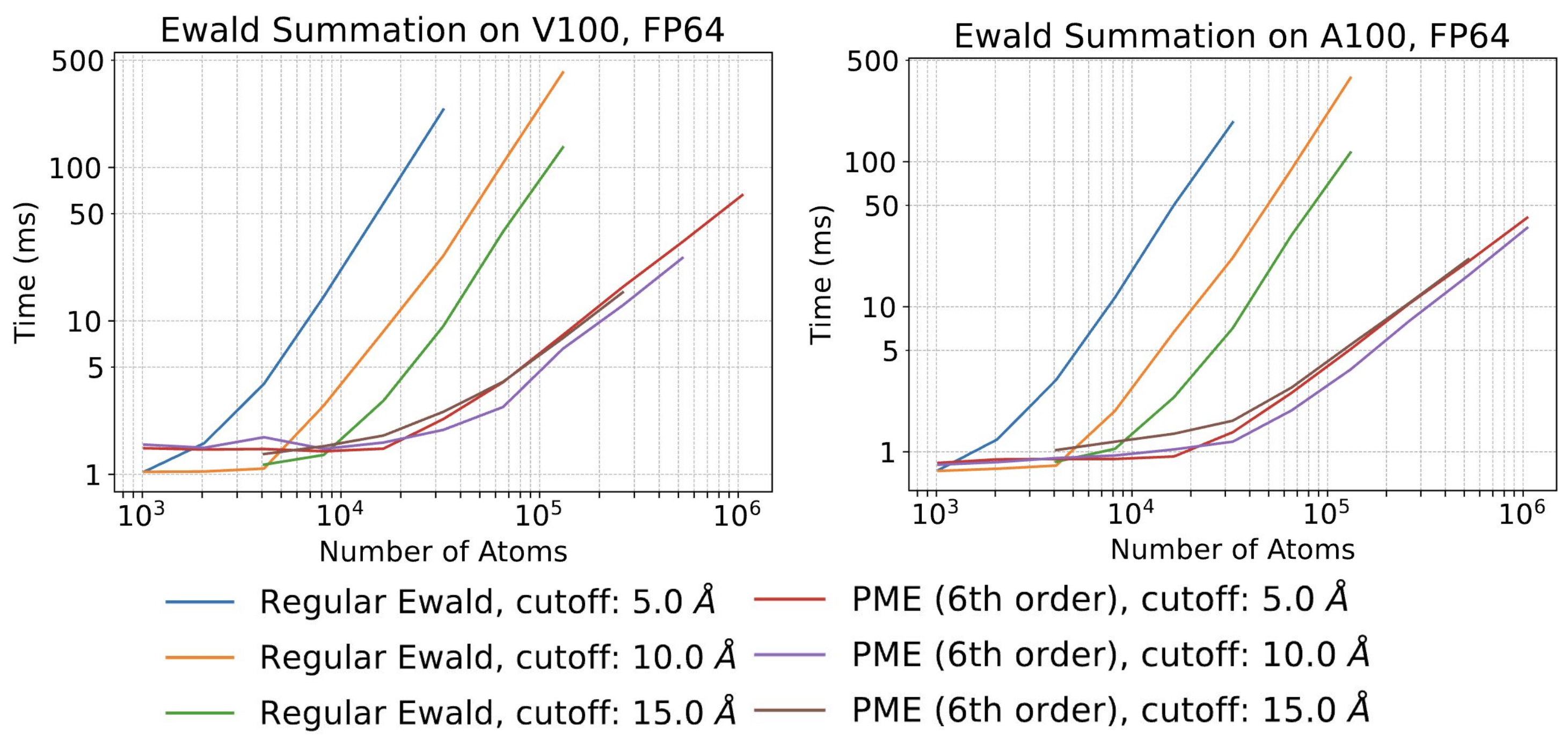}
    \caption{Detailed performance comparison of regular Ewald and PME (order-6) summation on A100 and V100 GPUs using various real space cutoffs.}
    \label{fig:ewald_perf_detailed}
\end{figure}

As the minimum image convention requires the smallest box dimension to be larger than twice the cutoff, a 15 \AA{} cutoff was too large for small systems. Additionally, a larger cutoff increases memory requirements for the real-space calculations. Due to these factors, certain data points in \cref{fig:ewald_perf_detailed} are missing.
\section{Hardware and Software}
\begin{table}[H]
\centering
\begin{tabular}{ll}
GPU          & CPU                                  \\ \hline
A100 (40 GB) & AMD EPYC 7702 (2 x 32 cores)         \\
V100 (32 GB)  & Intel Xeon E5-2660 v3 (2 x 10 cores)
\end{tabular}
\caption{Hardware used for the benchmarks}
\label{table:hardware}
\end{table}
\begin{table}[H]
\centering
\begin{tabular}{ll}
Software          & Version                                  \\ \hline
Python & 3.10         \\
PyTorch  & 2.5 \\
CUDA  & 12.4 \\
Triton & 3.1 \\
\end{tabular}
\caption{Software used for the benchmarks}
\label{table:software}
\end{table}

The GPU and CPU details used for the benchmarking are shown in \cref{table:hardware} and the related software versions are presented in \cref{table:software}.


\setcounter{enumiv}{0}
\bibliographySI{References}  
\bibliographystyleSI{unsrt}

\end{document}